\documentclass[iop,revtex4]{emulateapj}
\usepackage{enumerate}
\usepackage{graphicx}
\usepackage{amssymb}
\usepackage{epstopdf}
\usepackage{amsmath}
\usepackage[breaklinks]{hyperref}
\usepackage{ulem}
\newcommand{\teff}{\ensuremath{T_{\rm eff}}}
\newcommand{\logg}{\ensuremath{\log{g}}}

\newcommand{\feh}{[Fe/H]}

\newcommand{\msun}{\ensuremath{M_\sun}}

\newcommand{\rearth}{\ensuremath{R_{\earth}}}

\newcommand{\jband}{\textit{J}-band}
\newcommand{\hband}{\textit{H}-band}
\newcommand{\kband}{\textit{K}-band}

\newcommand{\jbande}{\textit{J}-band\ }
\newcommand{\hbande}{\textit{H}-band\ }
\newcommand{\kbande}{\textit{K}-band\ }

\newcommand{\eke}{\textit{Kepler}\ }
\newcommand{\dm}{\ensuremath{\Delta m}}

\begin{document}

\title{Assessing the Effect of Stellar Companions from High-Resolution Imaging of \eke Objects of Interest}

\author{
Lea~A.~Hirsch$^{1,\star}$,
David~R.~Ciardi$^2$,
Andrew~W.~Howard$^3$,
Mark~E.~Everett$^4$,
Elise~Furlan$^5$,
Mindy~Saylors$^{2,9}$,
Elliott~P.~Horch$^6$,
Steve~B.~Howell$^7$,
Johanna~Teske$^8$,
Geoffrey~W.~Marcy$^1$
}

\affil{$^1$University of California, Berkeley, 510 Campbell Hall, Astronomy Department, Berkeley, CA 94720}
\affil{$^2$NASA Exoplanet Science Institute Caltech-IPAC, Pasadena, CA 91125}
\affil{$^3$California Institute of Technology, Department of Astronomy; MC 249-17, Caltech, Pasadena, California 91125, USA}
\affil{$^4$National Optical Astronomy Observatory, 950 N. Cherry Ave, Tucson, AZ 85719}
\affil{$^5$Caltech-IPAC, Mail Code 100-22, Caltech, 1200 E. California Blvd., Pasadena, CA 91125}
\affil{$^6$Southern Connecticut State University, Dept of Physics, 501 Crescent St., New Haven, CT 06515}
\affil{$^7$NASA Ames Research Center, Moffett Field, CA 94035}
\affil{$^8$Carnegie DTM, 5241 Broad Branch Road, NW,Washington, DC 20015}
\affil{$^9$College of the Canyons, 26455 Rockwell Canyon Rd, Santa Clarita, CA 91355}
\altaffiltext{$\star$}{\small To whom correspondence should be addressed.  E-mail: lhirsch@berkeley.edu}

\slugcomment{Accepted to Astronomical Journal, January 13, 2017}


\begin{abstract}
\small We report on 176 close ($<2$'') stellar companions detected with high-resolution imaging near 170 hosts of \eke Objects of Interest. These \eke targets were prioritized for imaging follow-up based on the presence of small planets, so most of the KOIs in these systems (176 out of 204) have nominal radii $<6$  \rearth. Each KOI in our sample was observed in at least 2 filters with adaptive optics, speckle imaging, lucky imaging, or HST. Multi-filter photometry provides color information on the companions, allowing us to constrain their stellar properties and assess the probability that the companions are physically bound. We find that 60 -- 80\% of companions within 1'' are bound, and the bound fraction is $>90$\% for companions within 0.5'';  the bound fraction decreases with increasing angular separation. This picture is consistent with simulations of the binary and background stellar populations in the \eke field. We also reassess the planet radii in these systems, converting the observed differential magnitudes to a contamination in the \eke bandpass and calculating the planet radius correction factor, $X_{R} = R_p (true) / R_p (single)$. Under the assumption that planets in bound binaries are equally likely to orbit the primary or secondary, we find a mean radius correction factor for planets in stellar multiples of $X_R = 1.65$. If stellar multiplicity in the \eke field is similar to the solar neighborhood, then nearly half of all \eke planets may have radii underestimated by an average of 65\%, unless vetted using high resolution imaging or spectroscopy.  
\end{abstract}



\section{Introduction}
During the 4-year tenure of the \eke Mission, over 4000 planet candidates were identified and their radii estimated. The size of the \eke sample allows for detailed population statistics to determine the occurrence rates of planets of various sizes and orbital distances. Of primary interest is a determination of the occurrence rate of small planets, especially those with radii smaller than 1.6 \rearth, which could potentially be rocky \citep{Rogers2015, Marcy2014, Lopez2014}. The transition from ``non-rocky'' to ``rocky''  planets appears to be extremely sharp, making the detection of accurate planet radii critical for occurrence rate analyses \citep[e.g.][]{Rogers2015,Ciardi2015a}

Planet radius estimates from \eke light curves rely on the inferred properties of the planet host star. Due to the fairly low resolution and large pixel scale of the \eke detector, stellar companions are difficult to detect without ground-based follow-up observations. Higher-resolution imaging provides the most efficient method for detecting line-of-sight and bound companions within 1''. These companions, if present, can dilute the \eke light curves, preventing accurate assessment of the planet radius.

In the solar neighborhood, at least 46\% of sunlike stars have bound stellar companions \citep{Raghavan2010}, with the orbital separation distribution peaking at ~50 AU. Assuming the \eke field has similar multiplicity statistics, nearly half of the \eke target stars could have bound stellar companions falling within the same \eke pixel. 

Unknown close stellar companions will cause planetary radii to be underestimated, depending on the relative brightness of the binary components, and which star the planet is orbiting. For faint companions to planet hosts, the contamination from the secondary component may be limited to only a few percent; however, if the planet is determined to be orbiting the fainter secondary component, the planet radius estimate could be off by an order of magnitude \citep{Ciardi2015a}. Unfortunately, determining which star in a binary pair hosts the observed transiting planet is not straightforward, and requires extensive individualized statistical modeling \citep[e.g.][]{Barclay2015}. 

Neglecting to account for stellar multiplicity can have ramifications for the accuracy of occurrence studies of small planets. Since unknown stellar companions can cause larger planets to masquerade as ``small'' planets, stellar companions can artificially boost estimates of small planet occurrence by as much as 15 -- 20 \%  \citep{Ciardi2015a}. High resolution imaging follow-up is therefore extremely important for obtaining accurate measurements of planet radii.

Binary companions may also affect the formation and evolution of planets in a stellar system, and planets in binary systems are therefore an interesting sub-sample of the wider \eke sample. The full implications of stellar multiplicity on planet formation and evolution are still being explored. Previous studies \citep[e.g.][]{Wang2014a,Kraus2016} indicate that stellar multiplicity may be suppressed in planet-hosting \eke systems, compared to field stars in the solar neighborhood \citep{Raghavan2010}. This may indicate that planet formation is more rare in close binary systems than in single star systems. If this is the case, the problem of transit dilution from unknown stellar companions may impact fewer systems than expected based on field star multiplicity statistics. 

On the other hand, stellar multiple systems contain more than one star to host potentially discoverable planets, so their number may be enhanced in the KOI sample. Additionally, the augmented brightness of unresolved binaries relative to single stars of the same spectral type allows the flux-limited \eke survey to include stellar multiples at larger distances than corresponding single star systems. This might also augment the expected number of \eke systems in which flux dilution plays a role. The combined result of these various effects is complex, and more detailed statistical work on \eke binary statistics is still needed.

TRILEGAL galactic models with assumed multiplicity statistics were applied to the \eke field in order to determine the likelihood of chance alignments with background stars. These models indicate that companions observed within 1'' of a \eke Object of Interest (KOI) are highly likely to be bound companions, and therefore potentially play a dynamical role in the formation and evolution of the stellar system \citep{Horch2014a}. More specifically, companions within 0.4'' have a less than 10\% likelihood of being chance background alignments, and companions within 0.2'' have a nearly zero percent chance, based on the TRILEGAL galaxy models.

We study KOI host stars with detected close ($<$ 2'') companions on a case-by-case basis, in comparison to the statistical study performed by \cite{Horch2014a}. In~\autoref{sample}, we discuss the stellar sample of KOI hosts chosen for this study. 
In~\autoref{physical_association}, we describe the Dartmouth isochrone models used to determine whether a companion is physically associated based on color information, and the TRILEGAL models used to assess the background stellar population. In~\autoref{bound_companions}, we isolate a population of KOI hosts with bound companions and compare the bound fraction as a function of separation to the simulation results of \cite{Horch2014a}. Finally, in~\autoref{radius_corr}, we calculate radius correction factors for all of the planet candidates in our sample for use in future occurrence studies.

\section{Sample}
\label{sample}

\subsection{KOI Host Stars}
Our sample consists of 170 stellar hosts of KOIs observed with various high-resolution imaging campaigns. This sample was drawn from the overall sample of KOI stars observed with high-resolution imaging, described in the imaging compilation paper by \cite[][hereafter F17]{Furlan2017}.


We choose targets for this study by requiring that at least one companion was detected within 2'', and that the companion was detected in two or more filters, providing color information. We choose the 2'' separation limit to include all companions falling on the same \eke pixel as the primary KOI host star. For a complete list of detected companions, see F17. Because imaging campaigns typically targeted the hosts of small planets, our sample contains a high fraction of small planet hosts.

\begin{figure}[thb]
\centering
\includegraphics[width=0.5\textwidth]{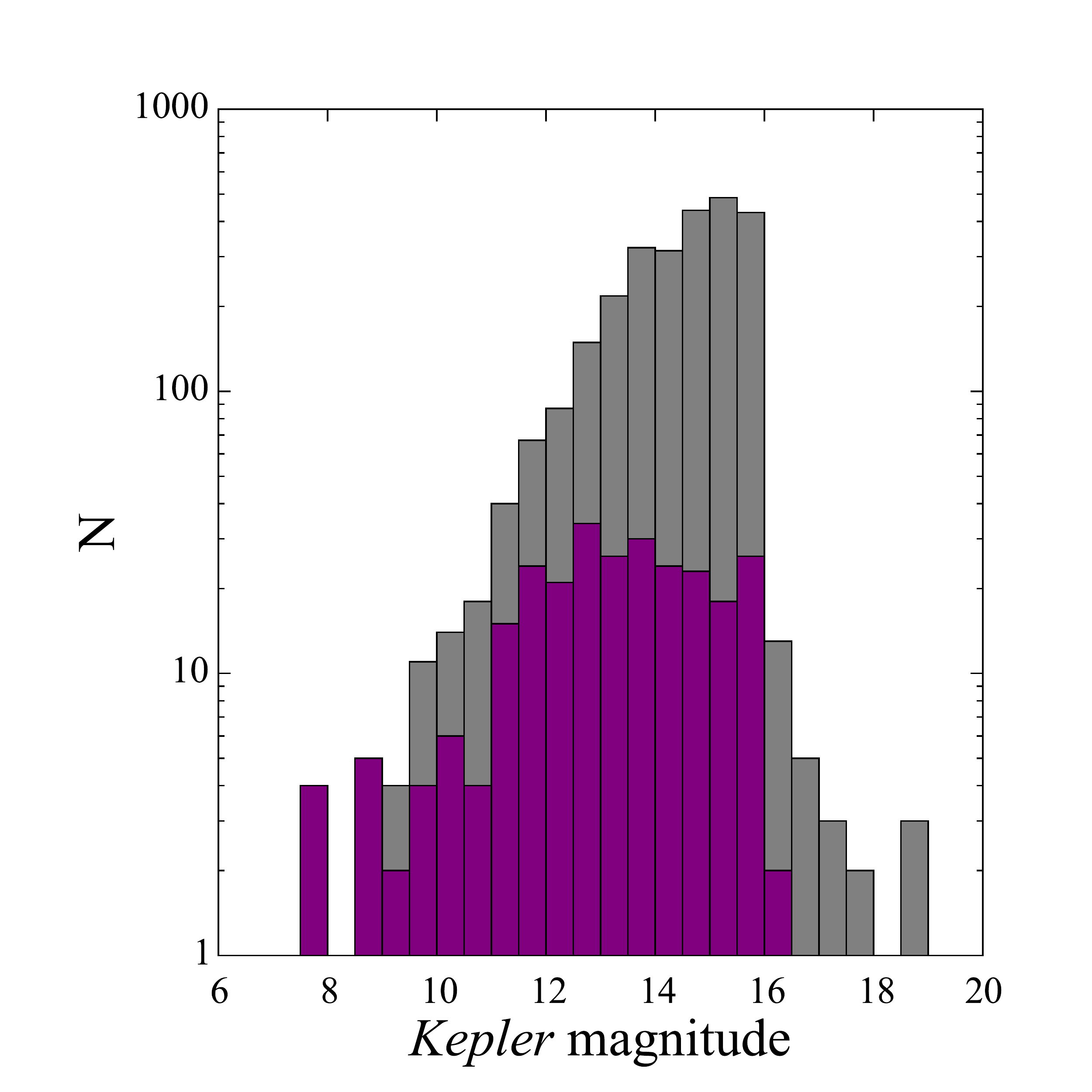}
\caption{Distribution of \eke magnitudes for stars in this study is plotted in purple. This is shown in comparison with the population of KOI hosts with any high-resolution follow-up (gray), including apparently single stars as well as stars with companions detected at separations $>2$'' or in only a single filter.
\label{fig:mag_histograms}}
\end{figure}

Figure \ref{fig:mag_histograms} is a histogram of the apparent \eke magnitudes of our sample KOI hosts in comparison with the full ensemble of KOI stars followed up with high-resolution imaging, described in F17. Although some efforts were made to image very faint \eke targets, none of our sample stars have \eke magnitudes fainter than 16. The majority of small \eke planets orbit faint host stars; however, this difficulty is currently remedied by the K2 and upcoming TESS missions.

For each stellar primary, the parameters \teff, \logg, and \feh\ are taken from \cite{Huber2014}. For many of the stars in our sample (71), these parameters come from spectroscopic or asteroseismic analysis performed by \cite{Huber2014,Chaplin2014,Huber2013,Ballard2013,Mann2013a,Batalha2013,Buchhave2012,Muirhead2012}. The spectroscopic and asteroseismic analyses provide stellar parameters largely unaffected by interstellar extinction toward the \eke field. These techniques are also largely unaffected by undetected close stellar companions, unless the companions' spectral lines contaminate the primary stellar spectrum to a significant degree. 

The remaining sample stars have stellar parameters calculated from photometry, reported in \cite{Huber2014}. Of these, all but 13 have properties revised by \cite{Pinsonneault2012}, \cite{Gaidos2013}, or \cite{Dressing2013}, using new models and data to improve on the original KIC-derived stellar parameters. For the purposes of our study, we do not re-analyze the stellar properties of any stars in our sample. Rather, we take the primary stellar properties from \cite{Huber2014} as given, and assess the characteristics of the companions detected around these stars. 

For the 13 stars with KIC stellar parameters, the stellar parameters are expected to be more uncertain than for stars with parameters derived from seismology, spectroscopy or improved photometric relationships. Additionally, in the case of systems with nearby companions, the photometry might be contaminated by the blended companion. The photometrically-derived stellar properties may therefore be biased. While we recognize this possible effect, it is beyond the scope of this paper to fully re-derive the stellar parameters for the primary stars, especially in light of the fact that many of the stars have parameters derived from spectroscopic and seismological observations.

\begin{figure*}[htb]
\centering
\includegraphics[width=\textwidth]{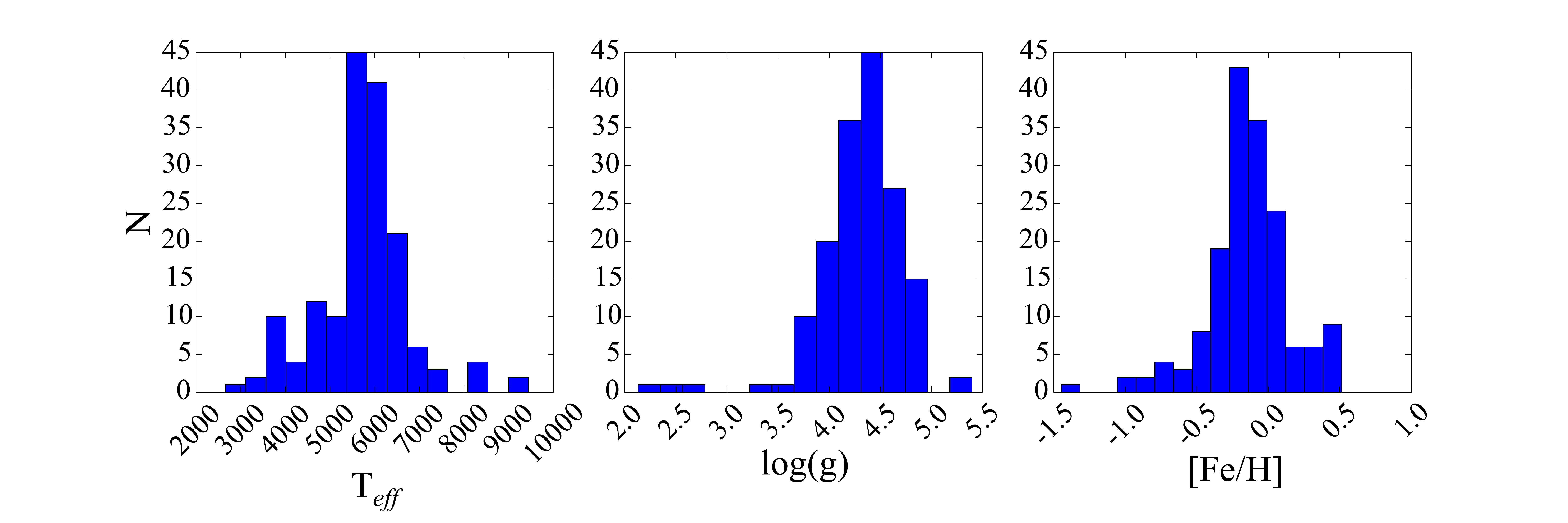}
\caption{Histograms of stellar parameters \teff, \logg, and \feh\ from \cite{Huber2014} for stars in our sample. The majority of our sample stars have \teff\ $< 7000$ K and \logg\ $> 4.0$. We include 5 stars that may be subgiants based on their reported \logg\ values (\logg\ $<  3.5$), as well as 5 hotter stars (\teff\ $> 8000$ K).
\label{fig:star_histograms}}
\end{figure*}

\begin{figure}[htb]
\centering
\includegraphics[width=0.5\textwidth]{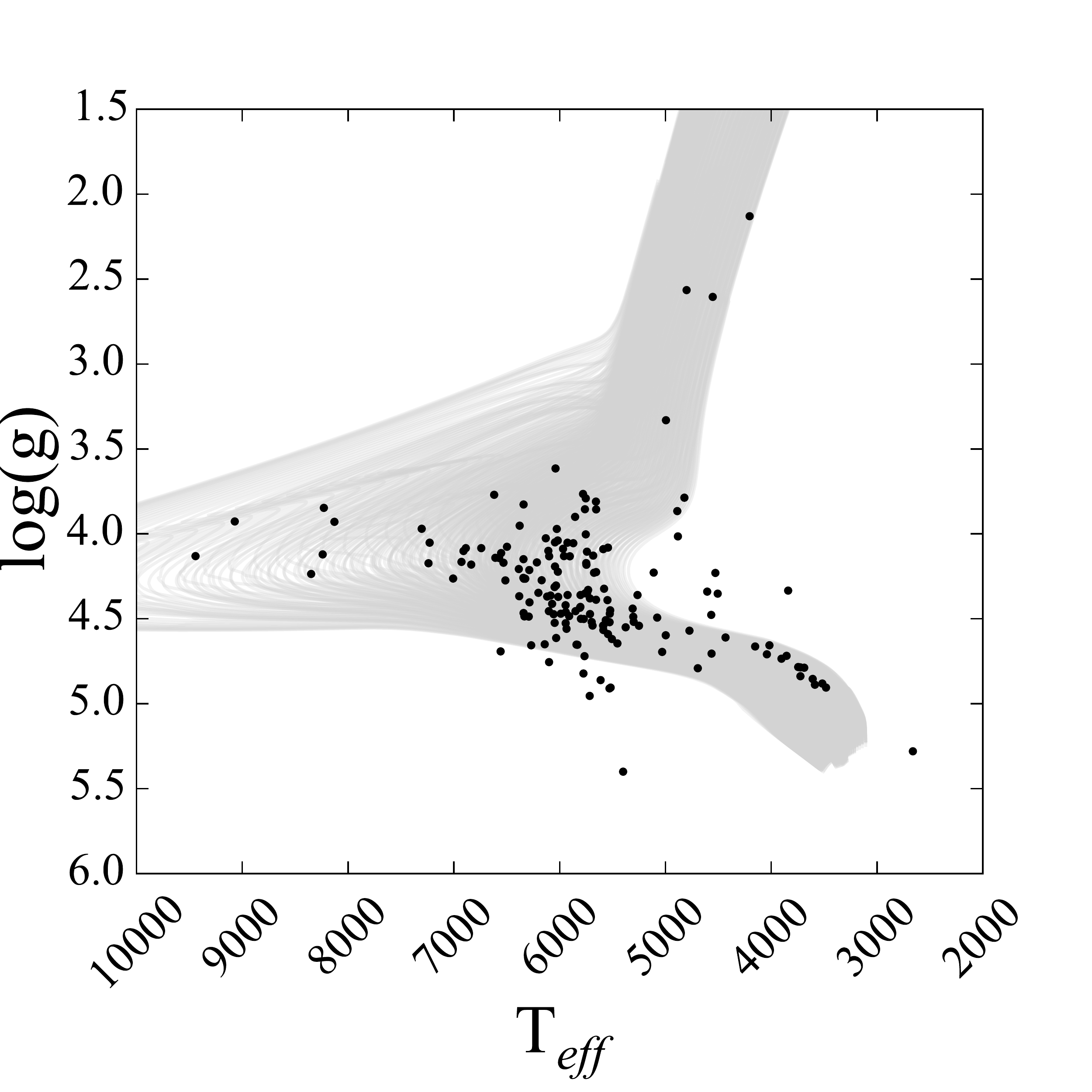}
\caption{HR Diagram of target KOI primary stars based on their stellar properties from \cite{Huber2014}. Dartmouth isochrones are displayed for context, spanning ranges in metallicity from -2.4 to +0.5 (in steps of 0.1 dex) and age from 1.0 to 15.0 Gyr (in steps of 1 Gyr). 
\label{fig:star_hr}}
\end{figure}

Figure \ref{fig:star_histograms} shows the distribution of stellar \teff, \logg, and \feh\ of our sample stars. The stars in this work are primarily G and K dwarfs, but the sample also includes 13 stars with \teff\ $< 4000$ K, and 6 hot stars with \teff\ $> 8000$K. We also include 4 evolved stars with \logg\ $< 3.5$ (see Figure \ref{fig:star_hr}). 

Our sample contains 110 confirmed planets in 72 systems. Of these, 15 planets are confirmed including analysis of high resolution imaging observations, allowing the authors to correct for the multiplicity of the host stars. These, we continue to treat as confirmed, and we do not reanalyze the planetary parameters. However, the remaining 57 systems were confirmed without the input of observations sensitive to stellar companions, so we do include these systems in our planet radius reanalysis. Eight of these systems were confirmed by nature of multiplicity by \cite{Rowe2014}, since transiting multi-planet systems have very low statistical probability of displaying false positive signals in addition to true planets. \cite{Morton2016} confirms 45 systems using statistical arguments where the possibility of the planets orbiting a secondary star within a stellar binary system was explicitly neglected. Finally, 2 systems were confirmed by \cite{VanEylen2015} and 3 were confirmed by \cite{Xie2013a} and \cite{Xie2014} using TTV analysis to confirm multi-planet systems. For these 57 systems, the high-resolution imaging follow-up data available to the public on the ExoFOP\footnote{\label{note1}https://exofop.ipac.caltech.edu/cfop.php} was not employed prior to confirmation, or the possibility of planets orbiting a stellar companion was not addressed. We therefore include these systems in our radius update analysis, while excluding confirmed systems with extensive analysis of high-resolution imaging data carried out by other groups.

Of the target systems, 76 have one or more planet candidate (PC), including many of the systems which also host confirmed planets. There are a total of 94 PCs in the sample. There are also 36 systems which host 38 false positive (FP) signals, including a few which also have planet candidates or confirmed planets. These systems were followed up with high-resolution imaging prior to receiving a false positive disposition. We assess whether the stellar companions of these systems are bound, but we do not include these transits in our analysis of planet radius updates in~\autoref{radius_corr}. 39 KOI stars in our sample host multi-planet systems, with 105 total planets or planet candidates among these systems.

It is important to note that the official \eke project used none of the ExoFOP\footnotemark[1] data (e.g., images) to assess candidates or FPs in the published planet candidate catalogues \citep[e.g.][]{Batalha2013,Burke2014,Rowe2015a,Mullally2015a}.

\subsection{Stellar Companions}
Table \ref{table:sample} documents the observations of companions to our KOI host star sample, compiled from the imaging compilation paper of F17. Columns (1) and (2) provide the KOI number and component of the companion. Columns (3) and (4) describe the companion's position relative to the KOI primary star, and column (5) describes the differential magnitude between the primary star and companion in the \eke bandpass, $\Delta Kp$, all taken from F17. Columns (6), (7), and (8) provide data leading to a determination of bound, uncertain, or unbound for each companion (9), which is explained in~\autoref{physical_association}.

F17 details the observations and measured differential magnitudes (\dm$= m_{2} - m_{1}$) for stars with high-resolution imaging, including our target systems. Each companion within 2'' must have at least two measured \dm\ values from the full set of filters used for follow-up observations, in order to be included in our sample. These filters include \jband, \hband, and \kbande from adaptive optics imaging from the Keck/NIRC2, Palomar/PHARO, Lick/IRCAL, and MMT/Aries instruments; $562\ nm$, $692\ nm$ and $880\ nm$ filters from the Differential Speckle Survey Instrument (DSSI) at the Gemini North and WIYN telescopes; $i$ and $z$ bands from the AstraLux lucky imaging campaign at the Calar Alto 2.2m telescope \citep{Lillo-Box2014}; and $LP600$ and $i$ bands from Palomar/RoboAO \citep{Law2014}. We also include seeing-limited observations in the $U$, $B$, and $V$ bands from the UBV survey \citep{Everett2012} and ``secure'' detections (noise probability $<$ 10\%) in the \jbande from the UKIRT \eke field survey.

In several of our higher-order multiple systems, only one component has detections in multiple filters, often due to field-of-view cutoffs or non-detections of very faint companions. We describe these additional companions in the notes section of~\autoref{table_refs}, and include them in the planet radius correction, but cannot include them in our physical association analysis. 

\begin{figure}[htb]
\centering
\includegraphics[width=0.5\textwidth]{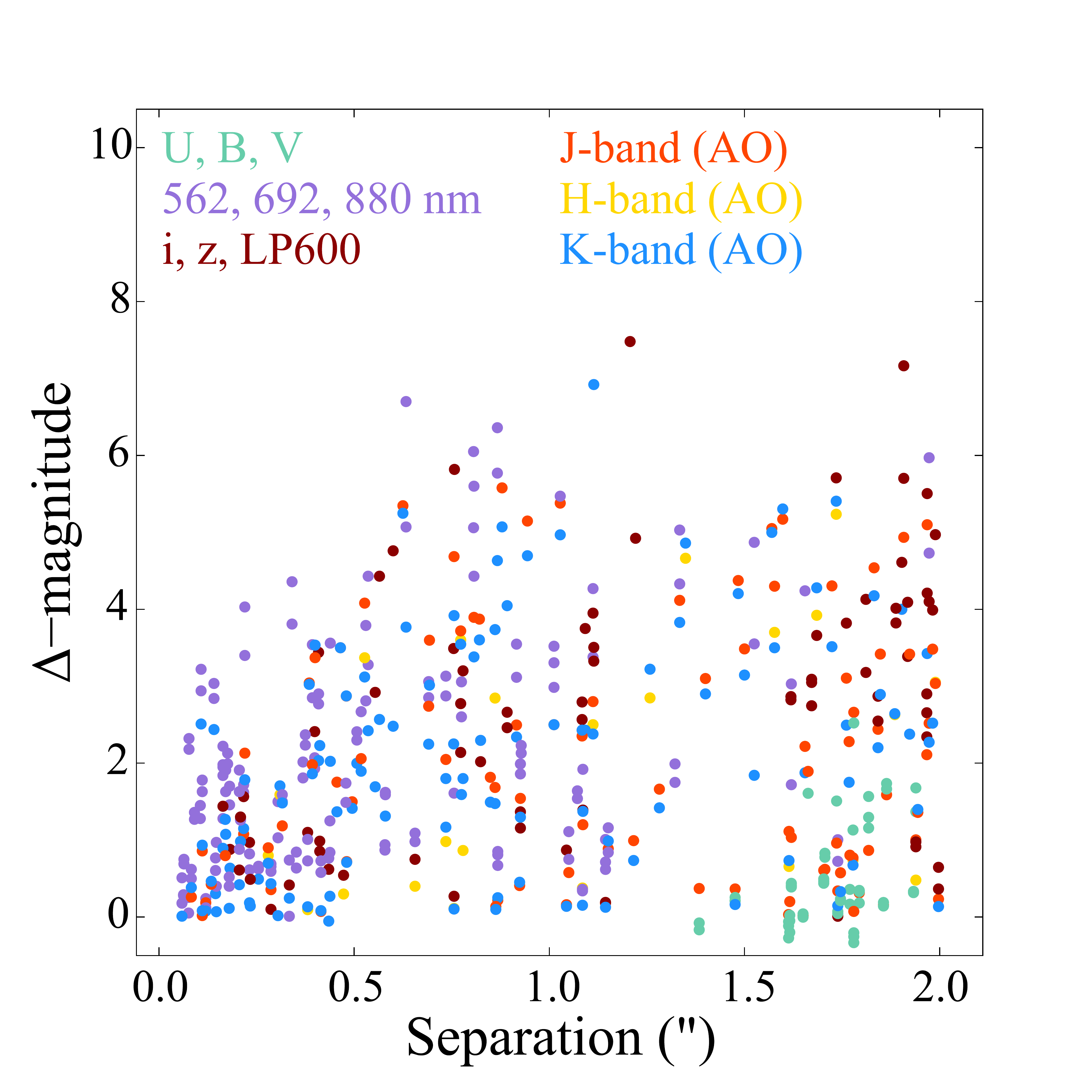}
\caption{Separation vs. \dm\ for each companion detected within 2'' of our sample KOI stars. Color denotes filter in which observations were made: $U$, $B$, and $V$ observations derive from the UBV seeing-limited survey. $562$, $692$, and $880\ nm$ observations derive from the DSSI instrument. $i$, $z$, and $LP600$ observations derive from RoboAO and AstraLux lucky imaging. \jband, \hband, and \kbande measurements come from adaptive optics imaging. Note that each companion has at least 2 measured \dm\ values, so each companion has 2 or more points on this plot, at the same angular separation.
\label{fig:comp_dist}}
\end{figure}

Figure \ref{fig:comp_dist} displays the separation versus \dm\ measured for each companion detected within 2'' of our sample KOI stars in DSSI, Astralux, RoboAO, UBV, and AO filters (see F17).

It is important to note that the term ``companion'' is used here to describe any star detected angularly nearby a KOI host star. It is not used to imply physical association. Instead, we will refer to physical binary or multiple systems as ``bound'' companions, and unphysical, line-of-sight alignments as ``unbound'' or ``background'' companions. 

\section{Determining Physical Association of the Companion Stars}
\label{physical_association}

Detection of a stellar companion near a KOI host star does not necessarily imply that the system is a physical binary. An assessment of the companion's stellar properties must be made to determine if the detected companion is physically bound, or simply a line-of-sight alignment of a background or foreground star. If the system is found to be a bound binary, then the companion may have played a dynamical role in the evolution of the planetary system; additionally, the planets may orbit either of the two stars in the system. If an unbound companion, only the flux dilution from the background source can be taken into account.

\begin{figure*}[htb] 
\centering
\includegraphics[trim={3cm 11cm 2cm 1cm},clip,width=0.7\textwidth]{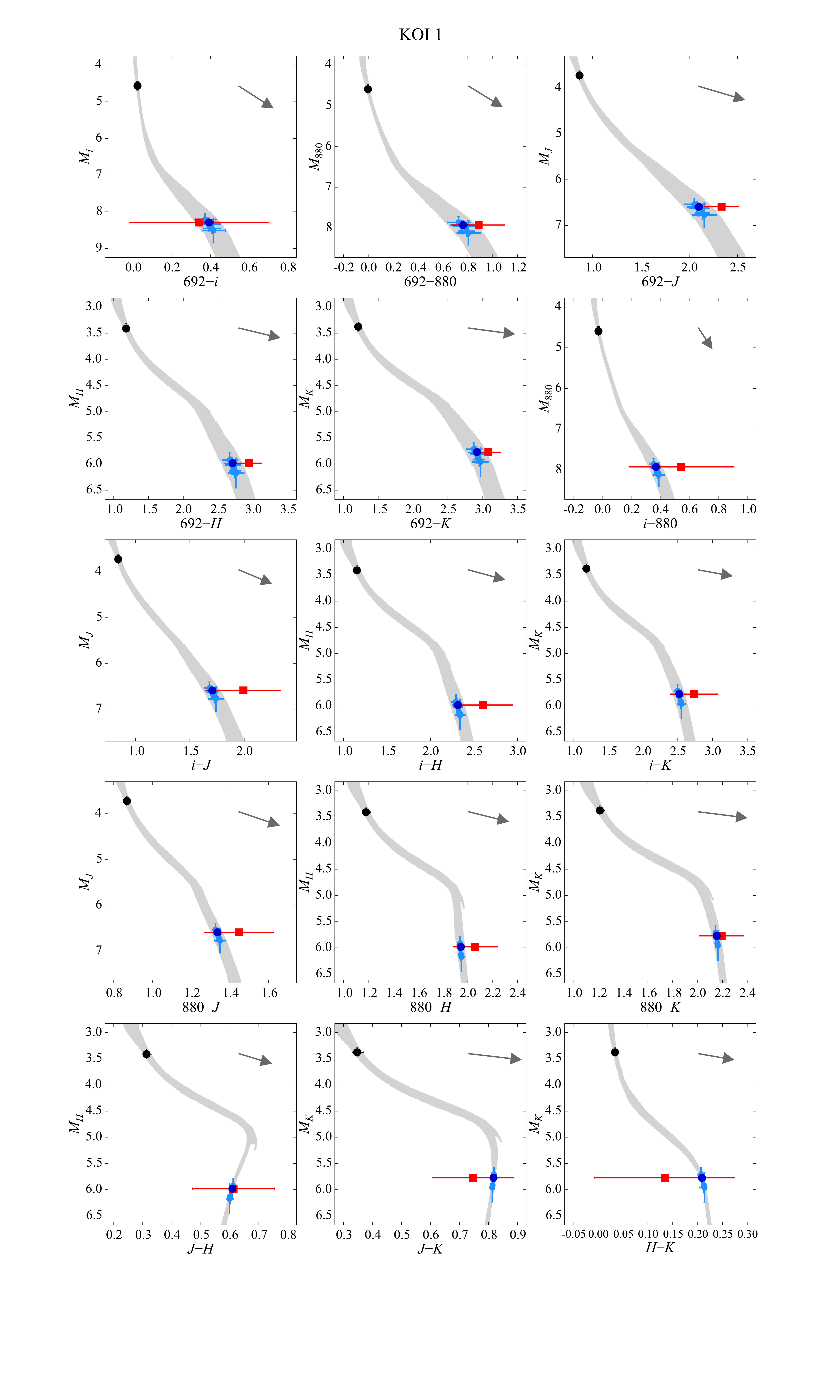}
\caption{This figure details the process we use to determine whether a companion (in this case, KOI 1 B) is bound. The isochrone model companion color (blue circle) is compared with the observed companion color (red square) to determine the likelihood that the companion is bound. Light blue points show the positions of each ``isochrone-shifted'' model estimate of the companion's position on the isochrone. The dark blue point is the weighted average model position, and is compared with the red observed color point. We plot an extinction vector corresponding to 1 magnitude of extinction in $V$-band, as a dark gray arrow in the upper right corner of each panel. This particular companion receives a designation of ``Bound'' because its modeled and observed colors agree to within 3$\sigma$, for all color pairings available.}
\label{fig:koi1}
\end{figure*}

\begin{figure*}[htb] 
\centering
\includegraphics[trim={3cm 11cm 2cm 1cm},clip,width=0.7\textwidth]{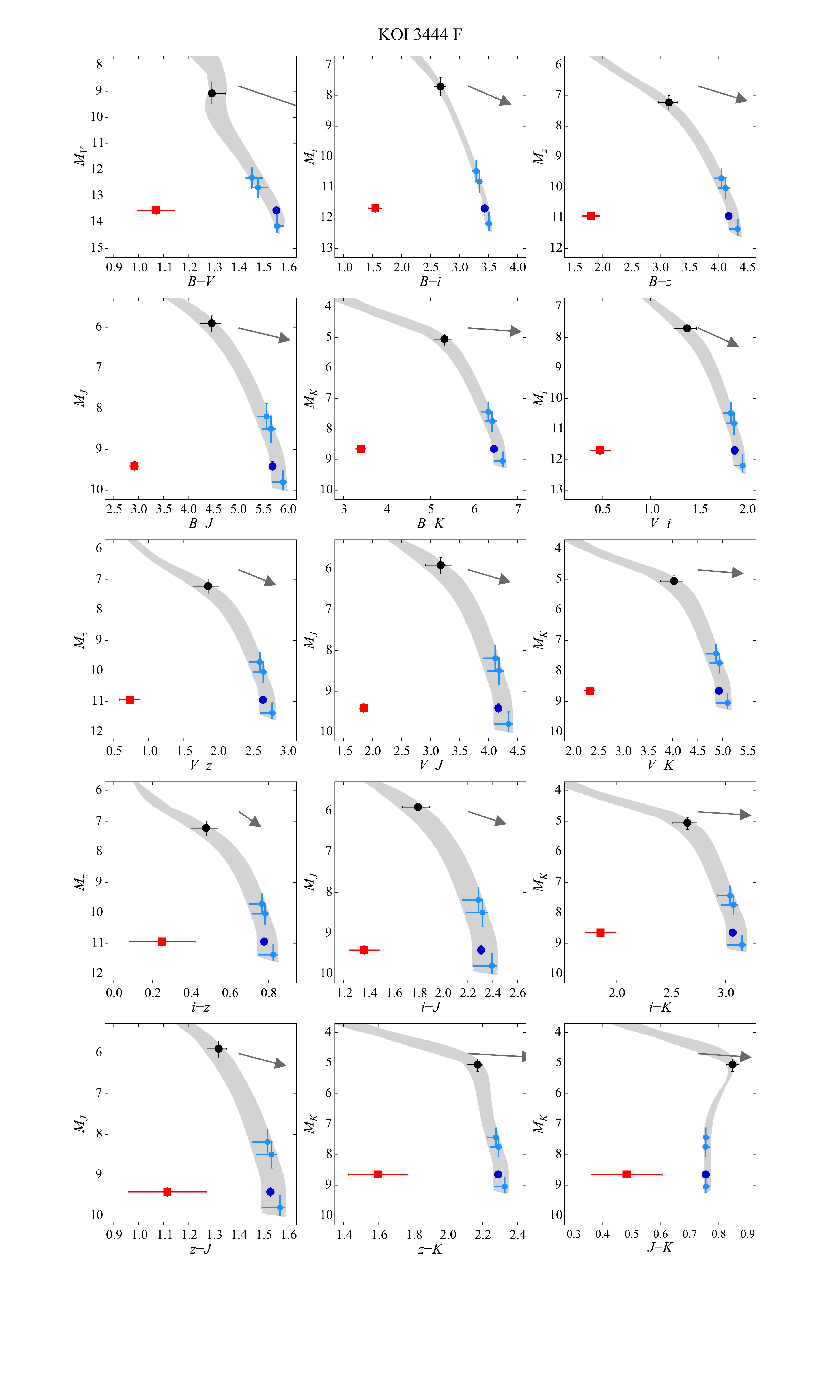}
\caption{For KOI 3444 F, the fifth stellar companion to KOI 3444, the isochrone model companion color (blue circle) is compared with the observed companion color (red square) to determine the likelihood that the companion is bound. Light blue points show the positions of each ``isochrone-shifted'' model estimate of the companion's position on the isochrone. The dark blue point is the weighted average model position, and is compared with the red observed color point. We plot an extinction vector corresponding to 1 magnitude of extinction in $V$-band, as a dark gray arrow in the upper left corner of each panel.This particular companion receives a designation of ``Unbound'' because its modeled and observed colors do not agree to within 3$\sigma$, for all color pairings available.}
\label{fig:koi3444F}
\end{figure*}

We use the Dartmouth Stellar Evolution models \citep{Dotter2008} to assess whether each detected companion is consistent with the isochrone of its respective primary star, by comparing the measured color to the color predicted for a model bound companion. This technique has been employed in previous works by \cite{Everett2015}, \cite{Teske2015}, and \cite{Wittrock2016}. We provide sample figures to elucidate this process in Figures \ref{fig:koi1} and \ref{fig:koi3444F}, choosing one unambiguously bound companion (KOI 1 B), and one unambiguously unbound companion (KOI 3444 F) to make the technique clear. Note that KOI 3444 F, at a separation of 3.5'', is not included in our 2'' sample, and was predicted to be unbound prior to analysis due to its larger angular separation from its primary KOI star. Not all companions have data available in as many colors as these two stars; many have only a single plot panel (one filter pair). A description of our technique and cutoffs follows:

\begin{enumerate}
\item{We use the Dartmouth models to produce a set of isochrones spanning ranges in metallicity from -2.5 to +0.5 (in steps of 0.02 dex) and age from 1 to 15 Gyr (in steps of 0.5 Gyr). We further interpolate these isochrones to achieve sampling in stellar mass between 0.1 and 4 \msun, with intervals no larger than 0.02 \msun. We also incorporate filter transmission curves for the DSSI filters, allowing absolute magnitudes to be predicted in these filters from the isochrone models.}

\item{We use the primary stellar parameters \teff, \logg, and \feh\ from \cite{Huber2014} to place the primary KOI host star on the isochrones and calculate a probability distribution for the primary star absolute magnitude in each of the filters in which that star was observed. Note that these distributions are subject to inaccuracy for the 13 stars with original KIC photometrically-derived stellar properties, which if bright, are likely to be more evolved than predicted by the original KIC.}

\item{To compute the {\bf observed} colors of the detected companions, we add the measured \dm\ values to the modeled absolute magnitudes of the primary stars, and subtract in pairs. This is particularly important for the stars detected with speckle imaging, since the DSSI filters are non-standard, and are not calibrated against more commonly-used filters. We therefore have no measured apparent magnitudes for the primary stars in these filters, from which we could deblend the component stars. Thus we use the isochrone magnitude of the primary to derive a magnitude of the companion from the observed \dm, then derive colors from the pairs of filters. For uniformity, we use this technique for all colors, even those including standard filters only (e.g. $J-K$), where observed apparent magnitudes could be used.}

\item{We compute isochrone models for a {\bf bound} stellar companion using each measured \dm. We propagate the primary star's probability distribution down the set of applicable isochrones by the measured \dm\ value. This provides an estimate of the stellar parameters of a bound stellar companion with the measured \dm. Each companion has $n$ ``isochrone-shifted'' models, where $n$ is the number of filters in which it is detected (at least 2 for each companion). We average these $n$ models to obtain a weighted-average companion probability distribution.}

\item{For each pair of applicable filters, we calculate the probable color of a hypothetical bound companion. We then compare this model color with the observed color of the companion to assess whether the companion is consistent with being bound. For a companion with \dm\ measurements in $n$ filters, we can calculate $\frac{n(n-1)}{2}$ colors. We use the weighted average companion model (which takes into account each measured \dm) to predict the model colors for every possible color pairing. We calculate the offset between model and observed color for each color pair, in units of the uncertainties in the measurements and models.

For a set of observations for a given star, the measured colors are not necessarily independent of each other (e.g., $J-K = (J-H) + (H-K)$).  However, to help mitigate the effects of using any one specific filter over the others, we have generated all possible colors for the filters available to assess the agreement between the observed and isochrone-model colors for each companion.}

\item{We take the average color offset between the models and observed colors, weighting by the measurement and model uncertainty in the two relevant filters for each color. This average color offset (as well as the weighted standard deviation of color offset) is provided in column (5) of Table \ref{table:sample}. If the average color offset is less than $3 \sigma$, we designate the companion as bound. If the average color offset is greater than $3 \sigma$, we designate the companion as unbound.}

\item{We isolate and reclassify systems where the designation is uncertain. First, we look for companions whose bound/unbound designation is reliant on a single very significant \dm\ measurement. We re-calculate the average color offset, removing the most significant observed color. If this procedure changes the conclusion, we reclassify the system as uncertain. Additional refinement of the \dm\ measurements is needed to determine whether the companion is bound or not.}

\item{Next, we look for systems with large color offset standard deviations, indicating that the conclusions from each color pair are mutually inconsistent for a single companion. If the conclusion can be changed by more than $0.5 \sigma$  by adding or subtracting the standard deviation from the average color offset (i.e. a seemingly unbound companion with $\langle \mathrm{offset}\rangle - \sigma_{\mathrm{offset}} < 2.5 \sigma$ or a seemingly bound companion with $\langle \mathrm{offset}\rangle + \sigma_{\mathrm{offset}} > 3.5 \sigma$), we reclassify the system as uncertain.}

\end{enumerate}

In Figures \ref{fig:koi1} and \ref{fig:koi3444F}, we plot a representative set of isochrones, with \feh\ within $\pm 1 \sigma$ of the primary star's input metallicity. In black, we plot the probability distribution of the primary star in color-magnitude space, based on the isochrone models for the primary star. Error bars indicate the $\pm 1 \sigma$ uncertainty contours.

The $n$ ``isochrone-shifted'' companion probability distributions are plotted in light blue, with the weighted average companion model in dark blue. Finally, we plot the measured companion color in red. The horizontal error bars on the observed color represent the measurement uncertainty in the \dm\ values used to calculate these colors. For $n$ filters, we have $\frac{n(n-1)}{2}$ panels in our plot, representing the number of color combinations possible with the set of filters available. Our analysis determines that KOI 1 B is consistent with being a bound companion, while KOI 3444 F is clearly unbound. 

The reader should note that this analysis makes the implicit assumption that the companion detected by each method is not itself an unresolved multiple. If a companion is itself an unresolved pair (i.e. a hierarchical multiple system), the observed \dm\ values would refer to the composite light from the companions, and the analysis may fail to accurately assess either companion star individually.

\subsection{Interstellar Extinction}
 Our approach makes the implicit assumption that the companion is subject to the same interstellar extinction as the primary star. We check for consistency between the model and observed companion colors under this assumption. 
 
Differential extinction between the primary and companion may cause the measured color to differ significantly from the color expected under the model assumptions, causing our analysis to produce a designation of unbound. Since the most likely scenario for producing this differential extinction is that the companion is more distant than the primary KOI star, and is therefore only a line-of-sight optical double, we would correctly conclude that the companion is unbound. 
  
One potential source for a false negative -- a bound companion that fails to satisfy our bound criteria -- is a bound companion with a significant dusty envelope or disk, causing distance-independent differential extinction of the companion. However, this scenario is highly unlikely, as the target stars are not expected to be young, so should not retain their natal dusty disks.
 
Likewise, a plausible source for false positives would be a background object that is attenuated and reddened until consistent with the isochrone of the primary star. To provide an indication of the direction and magnitude of the effect of extinction on background companions, we have plotted an extinction vector on each panel of Figures \ref{fig:koi1} and \ref{fig:koi3444F}. These vectors have lengths corresponding to one magnitude of extinction in the $V$-band. Typical models of interstellar extinction toward the \eke field assume 1 magnitude of extinction in $V$ for every kpc of distance in the galactic plane, with extinction diminishing as a function of galactic latitude with an e-folding scale-height of 150 pc \citep{Huber2014}. 


While these possibilities exist, we do not have enough data to make the assessment of whether the scenarios described apply to any of our systems. Spectroscopic data would provide more information about the intrinsic properties of the companion stars, and would allow a more conclusive assessment of whether the companions are indeed bound. However, spectroscopy is not available (and is largely precluded by magnitude limits and the closeness of the stellar primary) for the vast majority of the companions included in this sample. Instead, we look at the predicted population of background stars to determine the probability that each candidate bound companion is a line-of-sight alignment. We compare this to the probability of a solar-type primary hosting a bound stellar companion at the predicted mass ratio and separation.

\subsection{Comparison with Stellar Population Models}
\label{trilegal}
For the companions in our sample with colors consistent with the same isochrone as the primary (candidate bound companions), we additionally check the relative probabilities of the companions being bound versus background alignment, before applying the designation of bound. Specifically, we use the known multiplicity statistics of the Solar neighborhood from \cite{Raghavan2010} to determine the probability of a solar-type star having a companion within $\pm 3 \sigma$ of the model mass ratio from the isochrone models, and at a separation greater than the derived projected separation.

Based on the \cite{Raghavan2010} multiplicity statistics, we utilize a log-normal distribution for the probability distribution in orbital period, which we calculate based on the observed angular separation, the predicted stellar distances from the ExoFOP website\footnotemark[1], and the modeled stellar masses of both the primary and stellar companion. The distance estimates to each KOI star found on the ExoFOP website\footnotemark[1] derive from models calculated by \cite{Huber2014}. The uncertainties on these distance estimates are typically of order 15\%, and are uncorrected based on the discovery of a stellar companion. For KOIs with a bright stellar companion increasing the inferred brightness of the primary star, the true distance may be larger than the distance reported by the \cite{Huber2014} models. Greater distances would imply larger projected separations, and would therefore decrease the probability that the companion is bound.

To determine the probability of a companion within $\pm 3 \sigma$ of the modeled mass ratio, we assume a flat distribution in mass ratio q from values of zero to unity  \citep{Raghavan2010}. We integrate under the curve from the lower to upper bound, and apply this fraction to the overall fraction of sunlike stars with one or more stellar companions, 46\%. We then apply the fraction from integrating under the log-normal curve in orbital period, from the calculated period to infinity. We choose this upper bound to account for projection effects allowing long orbital periods to appear as smaller separations due to orbital orientation and phase.

We note that choosing the multiplicity statistics from \cite{Raghavan2010} may not be accurate for this sample, as previous work on the \eke field planet host stars has shown that stellar multiplicity may be affected by the presence of planets \citep{Wang2014,Wang2014a,Kraus2016}. However, the specifics of this multiplicity effect are not yet well-understood, so we default to using the known multiplicity statistics of the Solar neighborhood. 

We compare this derived bound probability to the probability of a chance alignment within the measured angular separation, and at the measured apparent \eke magnitude of the companion. We produce 9 galactic stellar population models using the online TRILEGAL galaxy population simulator \citep{Girardi2005} with galactic latitudes from $6^\circ$ to $22^\circ$, spanning our sample of KOIs, and areas of 1 square degree. For each KOI, we refer to the two galactic models with latitudes most similar to that of the KOI, and we interpolate to determine the background probability. We filter the results of each TRILEGAL model to contain only those stars with \eke magnitudes within $\pm 3 \sigma$ of the observed \eke magnitude of the stellar companion, since only stars with similar brightnesses and colors to the observed stellar companions would have been consistent with our initial check against the primary star's isochrone. We then calculate the stellar density of the field surrounding the KOI, and multiply by the area within the observed angular separation of the companion to determine the probability of a background interloper with the required brightness falling within the required area around the KOI.

We use only the \eke bandpass to filter the results of the galactic models, and still find background probabilities of $< 12$\% in all cases, and $< 2$\% in all but 3 cases. These very low background probabilities are expected, due to the very small angular region in which all of our observed companions reside. We find similar background probabilities for stellar companions designated as uncertain and unbound based on isochrone models, with no distinct difference among the 3 samples. These probabilities would be further reduced if we filtered the galactic models for consistency with all measured magnitudes or colors, instead of only looking at the \eke apparent magnitude. 

We calculate the ratio of bound probability to background probability for the candidate bound companions. In all but two cases, these ratios exceed unity, typically by one or more orders of magnitude. The median value of the probability ratio for stars in the bound category is 367.5. Only 11 stars have probability ratios less than 10, and only 2 (KOI 703 B and and KOI 6425 B) have ratios less than unity. We move these two companions to the uncertain category, and do not include them in further analysis of bound systems. In Table \ref{table:sample}, we provide the background probabilities, based on the TRILEGAL models, for all companions. We also provide the bound probabilities based on multiplicity statistics of the solar neighborhood \citep{Raghavan2010} for the companions consistent with being bound.

\section{Bound Companions to Kepler Planet Hosts}
\label{bound_companions}

\begin{figure*}[htb]
\centering
\includegraphics[width=0.8\textwidth]{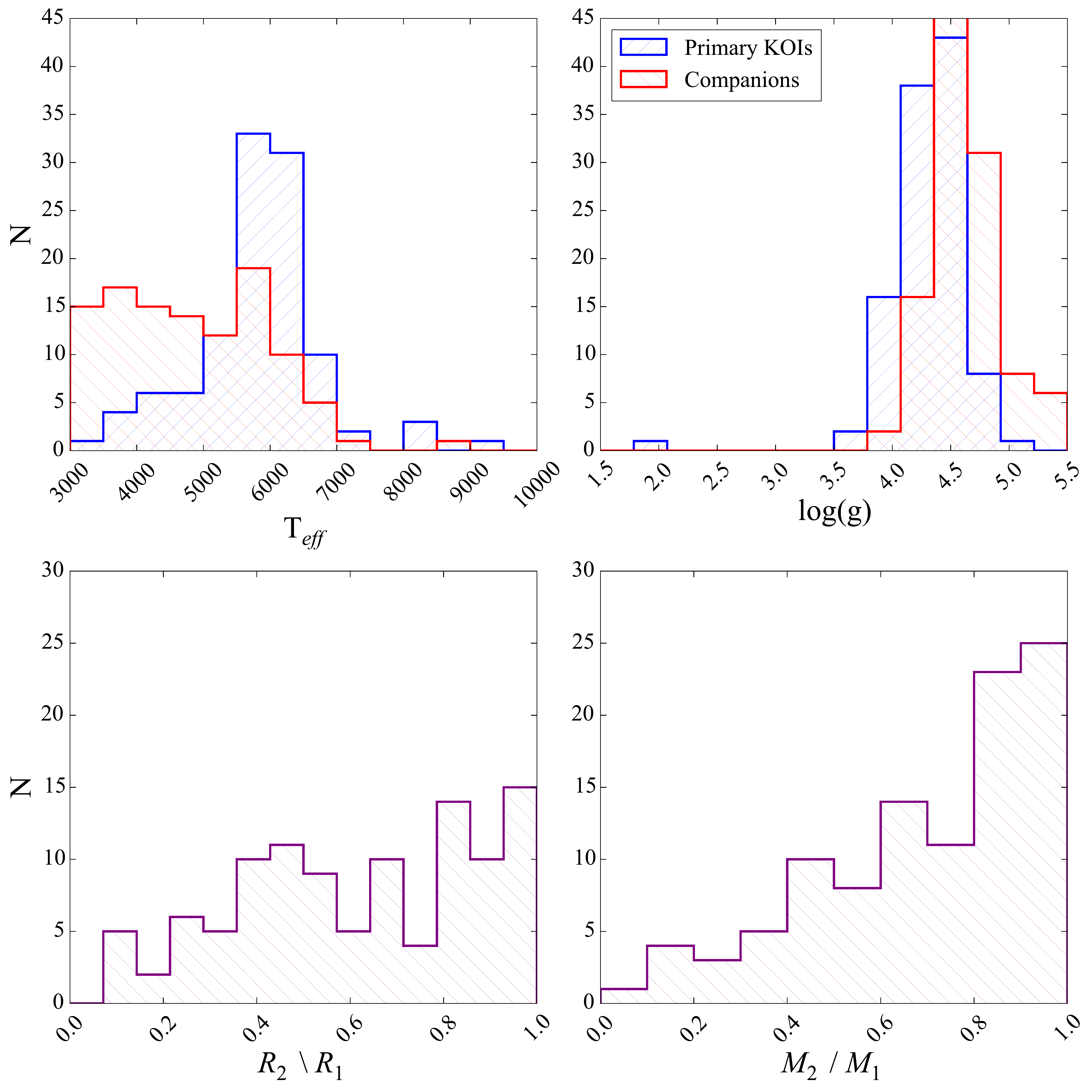}
\caption{Histogram of the stellar properties derived from isochrone models for the bound companions to KOI stars in our target sample are plotted in the diagonally hatched red histograms. The dark blue hatched histograms show the stellar properties of the primary KOI stars in each of these bound systems for reference. As expected, companion temperatures are skewed cool, with most companions having \teff \ $<$ 6000 K. These companions all appear to be main sequence stars, with $4.0 \leq \logg \leq 5.2$, assuming their host stars' stellar properties are accurate.  
\label{fig:bound_hists}}
\end{figure*}

\begin{figure}[hbt]
\centering
\includegraphics[width=0.5\textwidth]{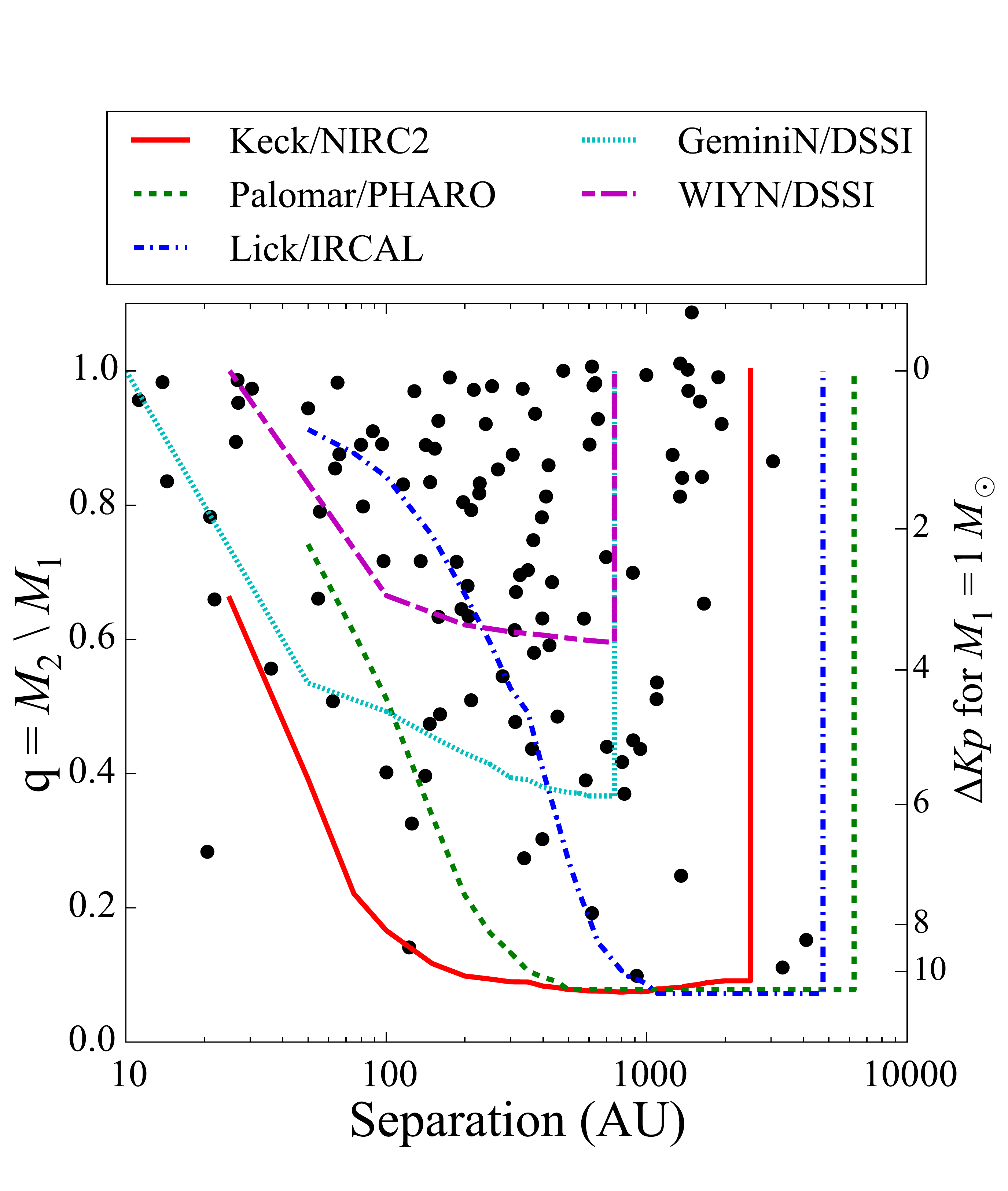}
\caption{Scatter plot of projected physical separation in AU versus mass fraction $q = \frac{M_2}{M_1}$ for bound companions. Each companion represents one data point on this plot. Projected separation is calculated from angular separation measured from imaging data and the distance estimate taken from the ExoFOP website. Each system's mass ratio is determined based on the isochrone models for the primary and secondary star. For context, we also plot the detection limits for several of the observational techniques used for this project, under the assumption that the primary star is sun-like ($M_1 = 1.0 $ \msun) and at a distance of 500 pc. These values represent a typical primary star for our sample, with a median primary star temperature of 5780 K and median distance of 420 pc. Regions of detectability lie {\bf above} the colored lines, and will vary for primary stars of different masses, at different distances. We also plot the corresponding $\Delta Kp$ values for systems with $M_1 = 1.0 $ \msun\ on the right axis for reference. The sensitivity curve for each instrument is plotted against the mass-ratio, not $\Delta Kp$, as each technique is carried out in a different filter, none of which include the \eke bandpass.
\label{fig:q_v_sep}}
\end{figure}

Based on the isochrone analysis and comparison to background stellar population models, we highlight a population of KOI hosts with likely bound companions, listed in Table \ref{table:sample}. This group of planet-hosting stellar multiple systems is a valuable comparison sample for the wider \eke planet host population. 

We plot the distributions of stellar properties for the bound companions to our sample targets in Figure \ref{fig:bound_hists}. We also plot for reference the stellar parameters of the primary stars in these bound binary systems. The companion population has a lower average \teff\ and a higher average \logg\ than their primary stars, as expected for a coeval pair. The bottom panels of Figure \ref{fig:bound_hists} show the radius and mass ratios of the bound stellar companions. This information feeds directly into the planet radius correction factors, described in~\autoref{radius_corr}. Note that this distribution is subject to detection limits on much fainter companions, so the lower number of stellar companions at small radius and mass ratio likely reflects the difficulty in detecting high-contrast systems. Malmquist bias may also play a role, since systems with mass and radius ratios close to unity would be brighter, and therefore overrepresented in both the \eke sample and the subsample of stars with high-resolution imaging follow-up.

In Figure \ref{fig:q_v_sep}, we plot the projected physical separations and mass ratios of each bound multiple system, in order to demonstrate the sensitivity of high-resolution imaging campaigns to companions in physical space. We calculate projected separation using the distance estimates to each KOI star found on the ExoFOP website\footnotemark[1]. The uncertainties on these distance estimates are typically $15 - 20$\%. Since the stellar companions to all of these KOI stars may increase the apparent brightness, stars with companions may be more distant than the models indicate, and the true projected separations may therefore be correspondingly larger. Mass ratios are calculated based on the isochrone models for each component of the system.

For reference, we provide characteristic detection limits for each of several techniques used to find the stellar companions here. We include AO observations from Keck/NIRC2, Palomar/PHARO, and Lick/IRCAL, as well as DSSI speckle observations from the Gemini North and WIYN Telescopes, taken from F17. We convert these detection limits from angular separation and \dm\ to projected separation and mass ratio by assuming the primary star is a solar-mass star at a distance of 500 pc. These are typical values for the primary stars in our sample, which have a median temperature of 5780 Kelvin and a median distance of 420 pc. Note that the exact detection limits will differ for each KOI system based on its primary star mass and distance.

In the future, the distribution of stellar systems in this diagram could be compared to a theoretical distribution, based on known properties of solar neighborhood binaries \citep[e.g.][]{Raghavan2010}, typical stellar properties and distances to stars in the \eke field, and the detection limits of each of the techniques used here. This comparison would allow us to test whether the multiplicity of \eke planet hosts is qualitatively similar to typical field binaries. However, this analysis is outside the scope of this paper, which is primarily focused on determining whether the companions discovered by high-resolution imaging are bound or not. For an ideal statistical study of stellar multiplicity as it relates to planet occurrence, a more careful selection of a target stellar sample, as well as a control sample, is required.

\cite{Horch2014a} simulate the binary and background stellar populations for the \eke field, and conclude that within the detection limits of the DSSI instrument at either the WIYN 3.5-m or Gemini North Telescopes, the vast majority of sub-arcsecond companions detected with high-resolution imaging should be bound, rather than line-of-sight companions. The bound fraction versus separation curve depends on the sensitivity of the instrument to high-contrast companions, and therefore can differ for each instrument and telescope. 

\begin{figure*}[htb]
\centering
\includegraphics[width=0.8\textwidth]{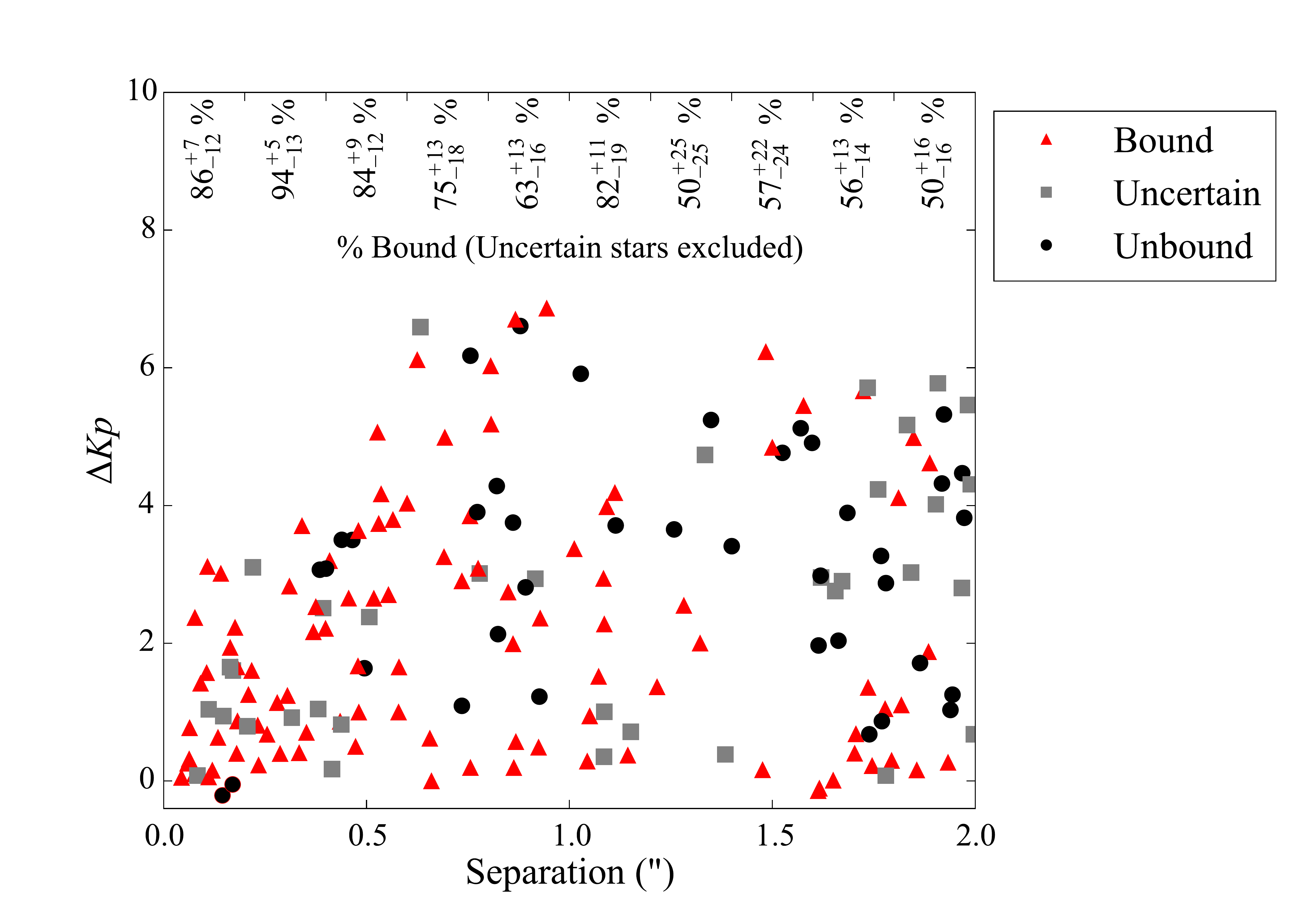}
\caption{Separation vs. $\Delta Kp$ for each companion detected within 2'' of our sample KOI hosts, with color denoting the designation we assign each companion (bound, uncertain, and unbound). The text in the top portion of the plot lists the bound fraction in bins of separation with width 0.2''. Confidence intervals are calculated based on a Clopper-Pearson binomial distribution confidence interval.
\label{fig:bound_frac}}
\end{figure*}

We use our bound and unbound/uncertain populations to assess the fraction of bound companions as a function of angular separation. In Figure \ref{fig:bound_frac}, we plot the separation in arcseconds against the $\Delta Kp$, the \dm\ in the \eke bandpass, for each companion in our sample, indicating with symbols which companions are bound, uncertain, and unbound. $\Delta Kp$ values are taken from F17 for unbound and uncertain stars, and from the isochrone models for bound stars. Visually, it is clear that companions within 1'' are most likely to be bound rather than either unbound or uncertain, while companions at 2'' are approximately equally likely to be bound or unbound. We list the fraction of companions found to be bound, in bins of 0.2'', in  Figure \ref{fig:bound_frac}. For the reported fractions, we exclude uncertain companions. 

We choose bins of this size in direct comparison with similar plots from \cite{Horch2014a}. The goal of this comparison is to assess the theoretical results with observational data. Our observational results confirm the model results from that study, which indicate that companions within 1'' are likely to be bound companions.

We do find 3 anomalous unbound companions at very close separations: KOI 270 B, KOI 4986 B, and KOI 5971 B. Only the first two of these appear in our plot, since KOI 5971 has insufficient data to calculate the $\Delta Kp$ value for the companion. We suspect that these might be anomalous false negatives, since a very bright, very close companion may skew the initial predicted stellar parameters of the primary KOI stars, which are used as the basis for our analysis. If the assumed stellar parameters of the primary star are incorrect, then the analysis may produce incorrect results. 

Indeed, two of these very close but apparently unbound companions (KOI 4986 B and KOI 5971 B) have stellar parameters estimated using KIC photometry by \cite{Pinsonneault2012}. Without knowledge of a very close, fairly bright stellar companion, the KIC photometry is blended and may produce inaccurate stellar properties. KOI 270, on the other hand, has stellar parameters from asteroseismology \citep{Huber2013}. Additional data are needed to resolve the anomaly of three such close, bright, but apparently unbound companions. For example, spectroscopic analysis might revise the stellar parameters calculated for the primary KOI host star, and repeating our analysis might then produce different results.

Between 0.2'' and 1.2'', our bound fraction results agree with those predicted by \cite{Horch2014a}, falling in between the prediction for the Gemini North and WIYN 3.5-m telescopes. At 0.0 -- 0.2'', our bound fraction is lower than predicted, mainly due to the three very close, unbound companions described above. If these three companions are removed, the bound fraction is 100\%, in agreement with the expected bound fraction in this separation region. Outside of 1.2'', \cite{Horch2014a} can not provide a prediction for the fraction of detected companions that should be bound, since the field of view of the DSSI camera excludes this region. We find that the bound fraction between 1.2'' and 2.0'' levels off, approaching approximately 55 \% of companions. At some separation, we expect this fraction to begin to decline; however, this occurs beyond our sample range. 

\begin{figure}[htb]
\centering
\includegraphics[width=0.5\textwidth]{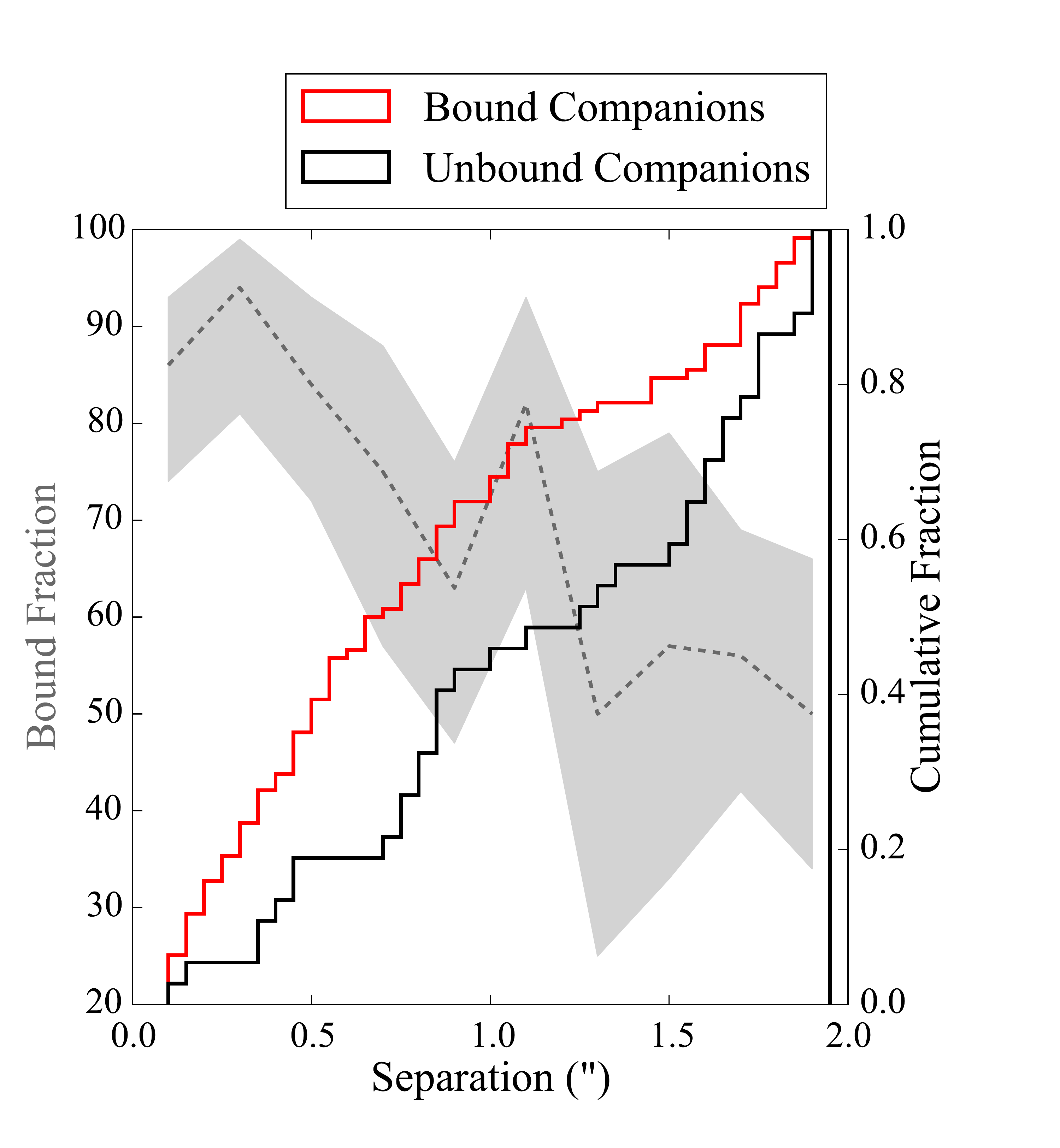}
\caption{On the left axis in gray shading, we plot the bound fraction as a function of separation, with $1\sigma$ confidence intervals. On the right axis, in heavy black and red lines, we plot the cumulative distributions of the separations of bound and unbound stellar companions, displaying statistically different populations. Bound companions are more tightly concentrated at close angular separations, while unbound companions are mostly more distant than 1''. 
\label{fig:bound_cumulative}}
\end{figure}

In Figure \ref{fig:bound_cumulative}, we plot the bound fraction with confidence intervals, listed in Figure \ref{fig:bound_frac}, in gray. We determine the 1$\sigma$ confidence intervals for the bound fractions using the Clopper-Pearson binomial distribution confidence interval. This method is based on the full binomial distribution.

We also plot the cumulative distributions of the separations of bound and unbound stellar companions on the right axis (in red and black, respectively), showing distinctly different distributions in separation for these two populations. Approximately 40\% of close bound companions are located within 0.5'' of their stellar primaries, and half of the detected bound companions  are located interior to 0.7''. On the other hand, the cumulative distribution places less than 50\% of unbound companions interior to 1.2''. 
 
 We perform a two-sample Kolmogorov-Smirnov test to determine whether the observed distributions of separations among the bound and unbound samples are consistent with being drawn from the same parent distribution. The null hypothesis that the two samples are drawn from the same parent distribution has a p-value of 0.002, indicating that the separation distribution for the bound companions is distinct from the distribution for unbound companions.

\section{Planet Radius Corrections}
\label{radius_corr}
Planetary radii for \eke planet candidates are calculated assuming the planet resides in a single stellar system \citep[e.g.][]{Batalha2013,Burke2014,Rowe2015a,Mullally2015a}. For KOI hosts with stellar companions unresolved in \eke photometry, however, the light from the stellar companion will dilute the transit signal, causing the planet to appear smaller than it actually is. 

For the purposes of the planet radius correction analysis, we whittle our KOI host sample down to a relevant subset. We first remove all KOIs with a disposition of False Positive, since these transit signals do not represent the presence of planets. We also remove several KOI systems with dispositions of Planet Candidate, but planetary radii larger than 2 $R_{J}$.

For confirmed planetary systems, we include only those systems without extensive and individual follow-up, including high-resolution imaging, available in the literature. For many of the confirmed systems in our sample, other research groups have discovered and accounted for the presence of a stellar companion in determining updated planet parameters. We do not repeat this analysis for these systems, since stellar multiplicity has already been addressed. 

This subsample does not exclude all confirmed planetary systems, since 57 of 72 confirmed systems in our sample were confirmed without an explicit analysis of the effects of the stellar companion. Of these, 8 achieved confirmed status by nature of their inclusion in multi-planet systems \citep{Rowe2014}. Five achieved confirmed status based on TTV analysis \citep{Xie2013a,Xie2014,VanEylen2015}. \cite{Morton2016} confirmed 45 systems statistically, but explicitly neglect the possibility that the planets might orbit a stellar companion. Since the planets in these 57 systems did not receive re-analysis including the full effects of the stellar companion, we must still account for the radius correction due to the stellar companion. For these confirmed systems, and for the planet candidate systems in our sample, we re-analyze the planetary radii to correct for the contamination of the transit signal by another star on the same \eke pixel.

For stellar multiple systems, the ratio of the true planet radius to the assumed single radius can be estimated as
\begin{equation}
X_R = \frac{R_p(\mathrm{true})}{R_p(\mathrm{single})} = \left(\frac{R_t}{R_1}\right) \sqrt{\frac{F_{tot}}{F_{t}}}
\label{eqn}
\end{equation}

where $R_t$ and $F_t$ are the radius and the \eke bandpass flux from the transited star, $F_{tot}$ is the \eke bandpass flux from all stars within the \eke aperture, and $R_1$ is the radius of the assumed-single primary KOI star \citep{Ciardi2015a,Ciardi2017}. This planet radius correction factor includes a component to correct for the dilution of the light from the stellar companion, as well as a component to correct for the assumption that the planet orbits the primary star, and thus $R_1$ is the relevant radius with which to calculate $R_p$. For any planets orbiting the KOI primary star,
\begin{equation}
\frac{R_t}{R_1} = 1
\end{equation}
since the primary star is also the planet host \citep{Ciardi2015a,Ciardi2017}. 

Note that Equation \ref{eqn} only provides an estimate of the radius correction factor for a given system. In order to more accurately assess the true planetary radii in a system with a close stellar companion, the original transit signal would need to be re-fit, in order to take limb-darkening  and a potentially updated impact parameter into account. In addition, knowledge of a stellar companion may change the interpretation of the primary star's stellar properties, particularly for those stars with photometrically-derived properties. These updates to stellar radius and \teff\ would also need to be addressed, on a case-by-case basis, in order to fully correct the planetary properties.

Without significant observational follow-up, we cannot fully assess which star in a stellar multiple system is the planet host. While an analysis of the position of the \eke raw image centroids has been used in some cases to assess the location of the transit host star \citep[e.g.][]{Barclay2015}, for the majority of the systems in our sample, the stellar companion is too close and the centroid position too poorly constrained to rule out either the primary or the companion as the planet host. Additional modeling may make this technique for determining the planet host possible for the widest stellar companions and the largest planets in the sample, but it is beyond the scope of this work. 

We therefore calculate $X_{R_n}$, the planet correction ratio assuming the planet host star is the $n^{th}$ component of the system, for each of our bound systems. In all cases but one, $n=1$ or $2$, where the planets may orbit either the primary or bound companion. We report these values in Table \ref{table:rad_corrs}. In the KOI 652 system, we detect 2 bound stellar companions, so we report $X_{R_3}$ in the table comments.

For unbound or uncertain systems, we are unable to perform the same analysis, since we do not have a good estimate of the stellar parameters of the line-of-sight companions. Due to unknown interstellar extinction, we cannot assume that the colors we measure are representative of the intrinsic colors of these stars, and we therefore cannot rely on color-\teff\ relations to estimate these stars' radii. We only calculate $X_{R_1}$, based on flux dilution, for these systems. However, it is still possible that the planets orbit the unrelated companion, rather than the KOI primary star. In this case, the planet radius correction factor may be significantly higher than the $X_{R_1}$ value we provide. We denote these systems in Table \ref{table:rad_corrs} with triple dots in the $X_{R_2}$ column. 

Converting from units of flux to magnitudes, we write the ratio of true to assumed single planet radius as
\begin{equation}
X_{R} = \frac{R_p(\mathrm{true})}{R_p(\mathrm{single})} = \left(\frac{R_t}{R_1}\right) \left(\sum_{n=1}^{N}{10^{-(\Delta m_n - \Delta m_t) / 2.5}} \right)^{1/2}
\end{equation}
where $\Delta m_n$ is the $\Delta Kp$ of the $n^{\mathrm{th}}$ component and N is the number of components in the system which fall on the \eke pixel, including the KOI primary star. $\Delta m_t$ is the $\Delta Kp$ of the transited star, and $\Delta m_t = 0.0$ if the primary star is the planet host.

Using the observed \dm\ values and colors of the companion stars, we calculate the degree of contamination due to the companion in the Kepler bandpass, $\Delta Kp$. For bound companions, we use the isochrone models to provide estimates of the \dm\ values in the \eke bandpass, using the weighted average of the absolute \eke magnitude from each of the ``isochrone-shifted'' models for the companion, as well as the absolute \eke magnitude of the primary provided by the initial isochrone models. 

For uncertain and unbound companions, we cannot use the isochrone models to accurately represent the stellar properties of the companion. Instead, we follow the conversions from the various measured \dm\ values to $\Delta Kp$ described in F17. They take a weighted average among all available methods for each target, excluding the $Kp-J$ and $Kp-K$ color-magnitude estimates for targets with both $J$ and $K$ measurements. This provides an estimate of the $\Delta Kp$ for each companion. We use this estimate for unbound and uncertain companions, whose isochrone models do not represent the background or foreground star observed.

We list the resulting $\Delta Kp$ values used for our analysis of the planet radius correction in columns (3) and (4) (if more than one companion lies within 2'' of the primary star) of Table \ref{table:rad_corrs}. Three stars in our sample (KOI 975, KOI 3214, and KOI 5971) do not have the correct \dm\ measurements to calculate $\Delta Kp$ as described. These stars were measured with filters for which we do not have relationships that would convert the measured \dm\ values to $\Delta Kp$. KOI 975 (Kepler-21) is analyzed in-depth and confirmed by \cite{Howell2012}, who include analysis of high-resolution imaging data, so we do not re-analyze the planet radii in this system. For the other 2 systems without calculated $\Delta Kp$ values, we cannot include a planet radius correction.

With known $\Delta Kp$ values, we calculate the radius correction factors assuming the planet orbits the KOI primary star, $X_{R_1}$. For bound companions we also calculate $X_{R_2}$, the radius correction factors assuming the closest bound stellar companion hosts the planets. The radius correction factors are listed in Table \ref{table:rad_corrs}. 

\begin{figure*}[htb] 
\centering
\includegraphics[trim={4.8cm 0 6.6cm 0},clip,width=0.9\textwidth]{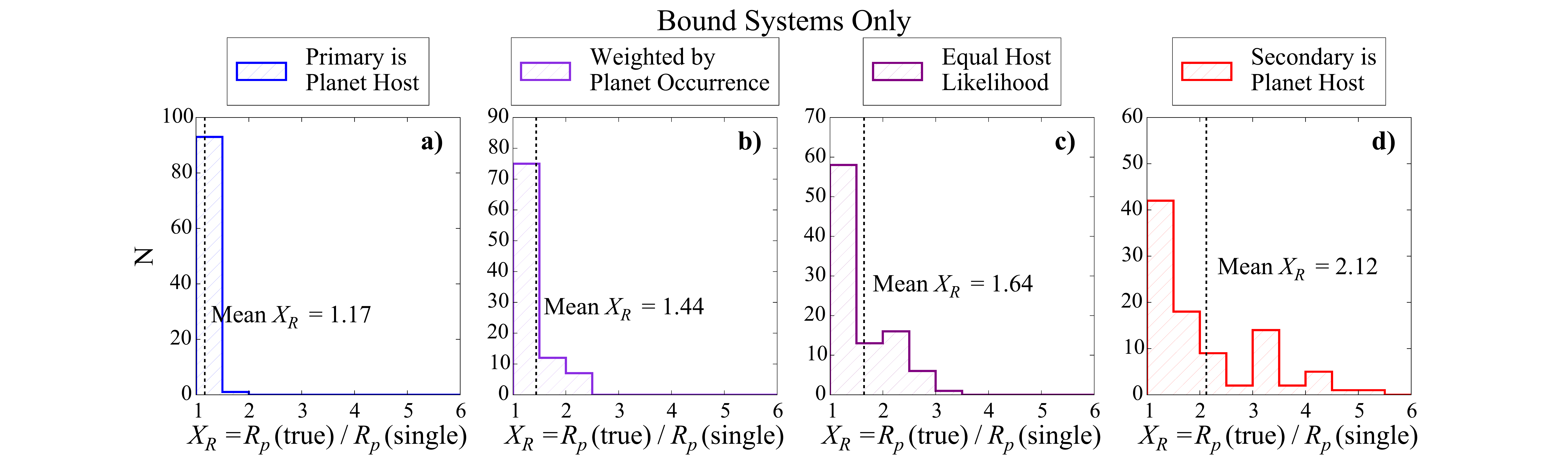}
\caption{Distribution of planet correction factors, $X_R$, for the bound systems in our sample. These include the dilution factor as well as correcting for the radius of the transited star. {\bf Panel a):} All planets assumed to orbit the primary KOI; $X_R = X_{R_1}$. {\bf Panel b):} $X_R$ calculated by weighted average, based on the relative occurrence rates of planets of various sizes. {\bf Panel c):} Planets equally likely to orbit primary or secondary star; $X_R = \langle X_R \rangle = 0.5\left( X_{R_1} + X_{R_2} \right)$. {\bf Panel d):} All planets assumed to orbit secondary star; $X_R = X_{R_2}$. In all cases, dilution is accounted for in the calculation of the correction factor.}
\label{fig:XR_dist_bound}
\end{figure*}

Figure \ref{fig:XR_dist_bound} shows the distributions of $X_{R}$ for the bound systems, under four scenarios: In panel a), we show the distribution of $X_{R_1}$, the dilution correction required to correct planet radii if all planets orbit the primary KOI stars. 

In panel b), we use the relative occurrence rates of planets of various sizes to assess the relative likelihood that the larger or smaller stellar component hosts the planets (making the planets smaller or larger, respectively). We use the binned occurrence rates as a function of planet radius, marginalized over planet orbital period, from \cite{Howard2012}. For a given planet, we calculate the radius assuming the planet orbits the primary star, and the larger radius assuming the planet orbits the secondary. We then compare the relative occurrence rates at these two planet radii, and weight our $X_R$ average by these occurrence values. 

The relative occurrence rates used for this analysis are marginalized across orbital period and stellar mass. This process is further complicated by a lack of understanding of the occurrence rates of planets in binary star systems, which may be very different from planets in single star systems. However, this simple analysis allows us to leverage previously-measured planet occurrence statistics and improve our estimate of the possible planet radius corrections.

In panel c), we assume the planets are equally likely to orbit the primary and secondary star, and take a simple average of the $X_{R_1}$ and $X_{R_2}$ values. In panel d), we show the distribution of $X_{R_2}$, the radius correction factors needed if all planets orbit the smaller secondary stars.

For KOI 652, which has two bound stellar companions within 2'', we only include $X_{R_2}$ in the plot. However, we provide an estimate of $X_{R_3}$ in the comments of Table \ref{table:rad_corrs}.

Under the assumption that planets in bound stellar multiple systems are equally likely to orbit either the primary or secondary star, a mean radius correction factor of $X_{R} = 1.65$ is found, indicating that planets in binary systems may have radii underestimated, on average, by 65\%.  The weighted average scenario may be more realistic, and provides a mean radius correction factor of $X_{R} = 1.44$.

\cite{Ciardi2015a} predict that the average radius correction factor for the \eke field should be $\langle X_{R} \rangle = 1.6$ for G dwarfs without radial velocity and high-resolution imaging vetting of KOI host stars. They assume the multiplicity of the \eke field is comparable to the solar neighborhood \citep{Raghavan2010}, and that planets are equally likely to orbit any component of a stellar multiple system. Comparable multiplicity to the solar neighborhood was shown to be a reasonable assumption by \cite{Horch2014a} based on high-resolution imaging, which is most sensitive to stellar companions at fairly large physical separations. However, there are indications that multiplicity may be suppressed at smaller separations \citep[e.g.][]{Wang2014,Wang2014a,Kraus2016}, and other effects such as Malmquist bias may also play a role in determining what fraction of \eke planet hosts have unknown stellar companions.

We would expect that our average measured $X_{R}$ for bound systems in our sample should be greater than the predicted $\langle X_{R} \rangle$ for the full \eke field, since we do not include any single stars ($X_{R} = 1.0$) in our analysis. If we make the simple approximation that our detected bound binary systems are representative of all binary systems in the \eke field (i.e. all \eke field binaries have a mean radius correction factor of $X_R = 1.65$, like our sample binaries), and assume that 46\% of all \eke stars are binaries (as the multiplicity statistics of the solar neighborhood indicate), then we would expect a mean radius correction factor for the full \eke field of $X_R = 1.3$, smaller, but roughly consistent with the predictions of \cite{Ciardi2015a}. Note that this simplistic analysis neglects unassociated background stars entirely, and a more complete analysis would also account for the variation in typical binary mass ratio (or \dm) as a function of separation.

\begin{figure*}[htb]  
\centering
\includegraphics[trim={4.8cm 0 3.5cm 0},clip,width=0.9\textwidth]{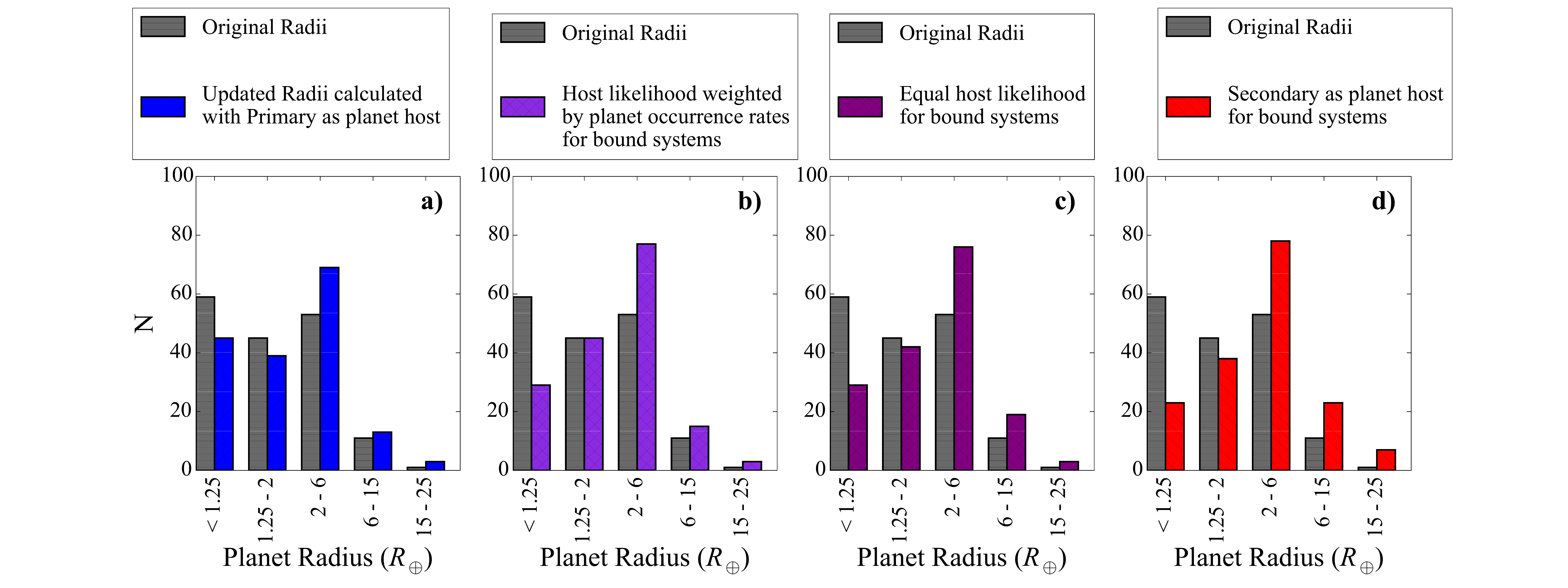}
\caption{Distributions of original \eke pipeline and updated planet radii for all planet candidates and confirmed planets without high-resolution imaging analysis in our sample. We divide the KOIs in our sample into bins representing planets of different types: Earth-sized planets ($<1.25$ \rearth), Super Earth-sized planets (1.25 -- 2.0 \rearth), Neptune-sized planets (2.0 -- 6.0 \rearth), Jupiter-sized planets (6.0 -- 15.0 \rearth), and larger planets ($>$ 15.0 \rearth). The gray bars plot the planet radius distribution of our sample under the assumption that the KOI hosts are single stars. The colored bars display updated planet radius distributions, under the same four scenarios as in Figure \ref{fig:XR_dist_bound}. In all cases, planets in unbound or uncertain systems are assumed to orbit the primary star, with $X_R = X_{R_1}$.}
\label{fig:radius_dist_all}
\end{figure*}

In Figure \ref{fig:radius_dist_all}, we plot the distribution of planet radii for KOIs in our sample, divided into specific, uneven bins corresponding to planet type: $<1.25$\rearth\ for Earth-sized planets; 1.25 -- 2.0\rearth\ for Super Earth-sized planets; 2.0 -- 6.0\rearth\ for Neptune-sized planets; 6.0 -- 15.0\rearth\ for Jupiter-sized planets; and $>15.0$\rearth\ for planets larger than Jupiter. In gray, we plot the number of planets in our sample in each of these bins, prior to radius correction. In color, we plot the distribution of corrected radii under the same four scenarios as in Figure \ref{fig:XR_dist_bound}, but including unbound and uncertain stellar systems as well as bound systems. In all cases, we only apply the dilution correction, $X_{R_1}$, to unbound and uncertain systems.

Figure \ref{fig:radius_dist_all} clearly shows that dividing planets into bins based on uncorrected planet radii can cause significant overestimates of the occurrence of small planets; in all four scenarios, the population of Earth-sized and Super Earth-sized planets in our sample decreases significantly when radii are corrected for the presence of stellar companions, while the number of Neptune-sized and Jupiter-sized planets increases. The amount by which the population sizes change depends upon assumptions about which star hosts the planets. This type of analysis, performed by \cite{Ciardi2015a} on a synthetic \eke population, predicts that the occurrence of small planets ($<1.6$\rearth) may be overestimated by 16\% unless stellar multiplicity is accounted for.

It is important to note that there is a corresponding effect acting in the opposite direction, potentially causing the occurrence of small planets to be underestimated. Specifically, the increased difficulty of finding small planets in systems with stellar companions means our sensitivity to small planets may have been overestimated. In truth, we are likely missing many small planets whose transits are heavily diluted by a stellar companion.

The relative scale of these two effects is difficult to quantify and is beyond the scope of this paper to calculate. However, we provide an estimate of the minimum detectable planet size for a typical binary system in our sample. KOI-1830 has an apparently bound stellar companion separated by 0.46'' from the primary KOI star, with a $\Delta Kp = 2.65\pm0.08$ mag, the median value of $\Delta Kp$ in our sample. KOI-1830 hosts two planets which were statistically validated by Morton et al. (2016), although the validation did not fully account for the stellar multiplicity. Prior to the discovery of the stellar companions, the best estimates of the planetary radii were $2.35\pm0.1$ and $3.65\pm0.15$ \rearth, with transit depths of 823 ppm and 2060 ppm. The orbital periods of the two planets are 13.2 and 198.7 days. Based upon our work, we find that the planets are underestimated by a factor of $X_{R_1}=1.042$ (planets orbit the primary) or $X_{R_2} = 2.409$ (planets orbit the secondary).

Based on the reported transit depths and S/N values, we estimate how small each of these planets could be before it became undetectable around each of the stellar components. SNR depends linearly on transit depth as:
\begin{equation}
\mathrm{SNR} = \frac{\delta}{\sigma_{CDPP}} \sqrt{\frac{n_{\mathrm{tr}} t_{\mathrm{dur}}}{3\ \mathrm{hr}}}
\end{equation} 
\citep{Howard2012}. 

We define a minimum transit detection threshold ($\delta_{threshold}$) such that the minimal signal-to-noise of a detected transit must be $\geq 7.1$ (as is the case for the Kepler pipeline).  For the two transiting planets (at their respective orbital periods), the minimum detectable transit depths are $\delta_{threshold} = $ 119 ppm and 346 ppm, respectively. Assuming no dilution by a stellar companion, these depths correspond to planetary radii of 0.89 \rearth\ and 1.5 \rearth.  However, taking into account the presence of the stellar companion, the minimum detectable planet sizes are 0.93 \rearth\ and 1.56 \rearth, if the prospective planets orbit the primary star, and 2.15 \rearth\ and 3.6 \rearth, if the prospective planets orbit the secondary star.  

Thus, while we might assume the data are sensitive to all planets in KOI-1830 that are larger than 1.5 \rearth\ with an orbital period of 198.7 days, planets as large as 3.6 \rearth\ might actually be present without being detected. Likewise, planets with sizes of 0.89 -- 2.15 \rearth\ might be undetectable at orbital periods of 13.2 days. 

The true frequency rates, and associated uncertainties, of planets as a function of planetary size must take into account the probabilities that the stellar systems contain more than one star.  Future efforts in this direction might allow us to assess the combined impacts of stellar multiplicity on planet occurrence statistics.

\section{Conclusions}
High resolution imaging follow-up is  essential for accurate estimation of transiting planet radii. It will prove especially useful for planets discovered by the K2 and TESS missions, due to their inclusion of closer, brighter stars than the \eke prime mission.

If stellar companions (both bound companions and chance line-of-sight background stars) cause significant underestimation of the \eke planet radii, this will certainly compromise calculations of planet occurrence, which divide the ensemble planet population into bins in planet size. Additionally, studies of the boundary between rocky and gas giant planets rely on comparing planet densities in bins of planet size. If both the densities and radii of planets are inaccurate due to a stellar companion, conclusions about the nature of small planets may be flawed.

We find that \eke planets in binary systems have radii underestimated by a mean radius correction factor of $\langle X_R \rangle = 1.65$, meaning their radii are underestimated, on average, by more than 50 \%, assuming equal likelihood of orbiting either component ($\langle X_R \rangle = 1.44$ if weighted by planet occurrence rates at different planet sizes). Expanding this to the full \eke sample, including single stars, we predict that the average radius correction factor for the full \eke field should be 1.3, roughly consistent with the simulation results from \cite{Ciardi2015a}. 

We confirm that very close stellar companions are highly likely to be bound, as predicted by \cite{Horch2014a}. We show that if a companion is detected within 2'', the star is likely bound ($\gtrsim 50$\%), and if detected within 1'', the star is almost certainly bound ($\gtrsim 80$\%).

Finally, we provide a demonstration of how uncorrected planet radii might bias a radius-based study of planet occurrence. For our sample, we find that correcting for stellar multiplicity or line-of-sight stellar companions significantly decreases the number of Earth-sized and Super Earth-sized planets, while increasing the number of Neptune- and Jupiter-sized planets in the sample. We also demonstrate the effect of stellar dilution making small planets that might otherwise be counted among Earth-sized and Super Earth-sized bins effectively undetectable, thus biasing small planet occurrence in the other direction.

The specific amount by which multiplicity biases the sample sizes of different types of planets depends heavily on assumptions about the multiplicity of the stars in the sample, as well as assumptions about which stars in a stellar multiple system host the planets. The best way to account for this problem is by using high-resolution imaging follow-up to rule out stellar companions whenever possible. This technique will be even more effective for the upcoming {\it TESS} and {\it K2} samples, which have much smaller average stellar distances than the \eke sample.

Although we do not perform a full statistical analysis of the multiplicity of the \eke planet hosts to assess how multiplicity affects planet formation and evolution in this work, we do provide a new sample of \eke planets in bound stellar multiple systems for this type of work in the future. A full analysis of the bias inherent in the selection of our sample would be required in order to make any statements about this subject.

\acknowledgements
We thank the anonymous referee for a thorough and detailed report and many helpful suggestions and comments.
This project was begun during and partially funded by the Infrared Processing and Analysis Center's Visiting Graduate Student Fellowship at the California Institute of Technology.
The authors would like to thank James Graham for useful discussions while working on this project. This research has made use of the NASA Exoplanet Archive and ExoFOP, which are operated by the California Institute of Technology, under contract with the National Aeronautics and Space Administration under the Exoplanet Exploration Program.

Dartmouth Stellar Evolution models \citep{Dotter2008}, TRILEGAL \citep{Girardi2005}

\bibliography{KOI_Bib_new}{}
\bibliographystyle{apj}

\clearpage
\LongTables
\begin{deluxetable}{ccccccccll}
\tabletypesize{\textwidth}
\tablecolumns{10}
\tablewidth{0pt}
\tabletypesize{\scriptsize}
\tablecaption{Observed Companions of KOI Host Sample
\label{table:sample}}
\tablehead{
\colhead{KOI} & \colhead{Comp} & \colhead{Separation} & \colhead{Position Angle} & \colhead{$\Delta Kp$} & \colhead{Color Offset} & \colhead{Background} & \colhead{Bound} & \colhead{Designation} & \colhead{Notes} \\
  &   & \colhead{('')} & \colhead{($^{\circ}$)} &  & \colhead{($\sigma$)} & \colhead{Prob. (\%)} & \colhead{Prob. (\%)} &  &  }
\startdata
1    & B  & $1.11\pm0.05$ & $136.20\pm1.14$ & $4.19\pm0.07$ & $0.60\pm0.37$  & 9.76e-03 & 2.66 &  Bound &  \\
5    & C  & $0.14\pm0.05$ & $304.30\pm2.16$ & $3.01\pm0.09$ & $1.36\pm0.98$  & 1.98e-04 & 7.47 &  Bound &  \tablenotemark{a} \\
13   & B  & $1.14\pm0.08$ & $279.70\pm4.28$ & $0.37\pm0.08$ & $1.71\pm1.23$  & 2.92e-04 & 2.01 &  Bound &  \\
14   & B  & $1.72\pm0.05$ & $273.50\pm1.04$ & $5.66\pm0.57$ & $0.50$  & 6.10e-01 & 3.23 &  Bound &  \\
42   & B  & $1.66\pm0.05$ & $35.80\pm2.09$  & $2.76\pm0.09$ & $2.99\pm0.77$  & 2.90e-03 &      &  Uncertain &  \\
68   & B  & $0.73\pm0.05$ & $256.50\pm2.12$ & $2.90\pm0.33$ & $1.27\pm0.43$  & 1.96e-02 & 4.44 &  Bound & \tablenotemark{b} \\
97   & B  & $1.90\pm0.05$ & $105.10\pm3.66$ & $4.02\pm0.08$ & $18.05\pm1.31$ & 1.06e-01 &      &  Uncertain &  \\
98   & B  & $0.29\pm0.05$ & $144.00\pm1.87$ & $0.40\pm0.07$ & $2.34\pm0.70$  & 1.09e-04 & 3.95 &  Bound &  \\
105  & B  & $0.86\pm0.06$ & $157.70\pm1.62$ & $3.75\pm0.81$ & $25.85$ & 5.20e-01 &      &  Unbound & \tablenotemark{c} \\
112  & B  & $0.11\pm0.05$ & $115.90\pm4.70$ & $1.04\pm0.08$ & $2.64\pm0.78$  & 7.46e-05 &      &  Uncertain &  \\
113  & B  & $0.17\pm0.05$ & $167.70\pm2.24$ & $1.60\pm0.15$ & $3.98\pm1.65$  & 5.30e-04 &      &  Uncertain & \tablenotemark{d} \\
118  & B  & $1.33\pm0.10$ & $213.80\pm1.16$ & $4.73\pm0.09$ & $3.03\pm0.53$  & 6.78e-02 &      &  Uncertain &  \\
119  & B  & $1.04\pm0.05$ & $119.30\pm1.37$ & $0.29\pm0.14$ & $1.39\pm1.20$  & 7.10e-03 & 1.21 &  Bound &  \\
120  & B  & $1.62\pm0.15$ & $129.20\pm1.43$ & $2.96\pm0.15$ & $3.27\pm2.95$  & 5.02e-02 &      &  Uncertain &  \\
141  & B  & $1.09\pm0.06$ & $12.80\pm1.66$  & $1.00\pm0.08$ & $17.86\pm2.95$ & 6.96e-03 &      &  Uncertain &  \\
174  & B  & $0.56\pm0.07$ & $76.90\pm4.02$  & $3.79\pm0.10$ & $0.71$  & 2.23e-02 & 2.64 &  Bound &  \\
177  & B  & $0.23\pm0.05$ & $215.80\pm2.99$ & $0.81\pm0.20$ & $0.68\pm0.13$  & 1.83e-03 & 5.48 &  Bound & \tablenotemark{e} \\
190  & B  & $0.18\pm0.05$ & $105.00\pm3.51$ & $1.66\pm0.46$ & $1.61$  & 2.47e-03 & 8.80 &  Bound &  \\
191  & B  & $1.67\pm0.05$ & $96.10\pm1.59$  & $2.90\pm0.02$ & $13.06\pm1.57$ & 8.20e-02 &      &  Uncertain &  \\
227  & B  & $0.30\pm0.05$ & $69.10\pm1.64$  & $1.24\pm0.14$ & $2.73$  & 1.59e-03 & 7.94 &  Bound & \tablenotemark{f} \\
258  & B  & $1.01\pm0.05$ & $72.80\pm1.09$  & $3.37\pm0.63$ & $1.83\pm1.55$  & 2.11e-02 & 8.15 &  Bound &  \\
258  & C  & $1.40\pm0.07$ & $73.90\pm1.41$  & $3.41\pm0.08$ & $4.85$  & 4.24e-03 &      &  Unbound &  \\
264  & B  & $0.47\pm0.05$ & $38.80\pm4.15$  & $3.50\pm0.08$ & $15.51$ & 2.16e-03 &      &  Unbound &  \\
268  & B  & $1.76\pm0.06$ & $267.70\pm3.10$ & $4.24\pm0.08$ & $0.99\pm0.48$  & 2.53e-02 &      &  Uncertain & \tablenotemark{g} \\
270  & B  & $0.14\pm0.06$ & $64.40\pm2.69$  & $-0.21\pm0.07$ & $3.37\pm0.51$ & 2.21e-05 &      &  Unbound &  \\
279  & B  & $0.92\pm0.05$ & $246.90\pm1.05$ & $2.94\pm0.07$ & $5.88\pm2.71$  & 7.01e-03 &      &  Uncertain &  \\
284  & B  & $0.87\pm0.05$ & $97.20\pm1.09$  & $0.57\pm0.14$ & $1.91\pm0.84$  & 1.73e-03 & 3.87 &  Bound &  \\
285  & B  & $1.48\pm0.05$ & $137.80\pm1.27$ & $6.23\pm0.11$ & $1.89$  & 1.44e-01 & 4.03 &  Bound & \tablenotemark{h} \\
287  & B  & $1.07\pm0.05$ & $30.50\pm1.00$  & $1.52\pm0.60$ & $1.26$  & 3.75e-02 & 7.73 &  Bound &  \\
292  & B  & $0.39\pm0.05$ & $122.00\pm1.20$ & $3.07\pm0.08$ & $8.79$  & 1.82e-03 &      &  Unbound &  \\
294  & B  & $1.60\pm0.05$ & $249.90\pm1.02$ & $4.91\pm0.07$ & $10.66$ & 1.28e-01 &      &  Unbound &  \\
298  & B  & $2.00\pm0.06$ & $272.80\pm1.05$ & $0.68\pm0.01$ & $15.85\pm5.09$ & 1.71e-03 &      &  Uncertain & \tablenotemark{i} \\
300  & B  & $0.16\pm0.05$ & $317.60\pm1.31$ & $1.94\pm0.50$ & $0.11$  & 1.70e-03 & 12.13 & Bound &  \\
307  & B  & $0.08\pm0.05$ & $244.80\pm4.90$ & $0.10\pm0.18$ & $0.90$  & 5.25e-05 & 5.56 &  Bound &  \\
335  & B  & $1.86\pm0.07$ & $117.20\pm1.01$ & $0.16\pm0.40$ & $0.67\pm0.29$  & 1.02e-01 & 3.47 &  Bound &  \\
356  & B  & $0.55\pm0.05$ & $217.60\pm1.19$ & $2.70\pm1.24$ & $1.08$  & 2.28e-01 & 12.75 & Bound &  \\
364  & B  & $0.09\pm0.06$ & $100.00\pm3.79$ & $1.42\pm0.12$ & $0.05$  & 1.77e-05 & 12.42 & Bound &  \\
378  & B  & $1.79\pm0.05$ & $317.30\pm1.00$ & $0.30\pm0.41$ & $0.60\pm0.35$  & 1.31e-01 & 5.58 &  Bound & \tablenotemark{j} \\
379  & B  & $1.09\pm0.06$ & $303.40\pm1.57$ & $2.28\pm0.38$ & $1.60$  & 1.56e-01 & 6.96 &  Bound &  \\
379  & C  & $1.94\pm0.05$ & $80.20\pm1.05$  & $1.03\pm0.00$ & $7.61\pm2.33$  & 2.09e-03 &      &  Unbound &  \\
387  & C  & $0.89\pm0.06$ & $350.50\pm1.10$ & $2.81\pm0.02$ & $4.96$  & 3.67e-03 &      &  Unbound & \tablenotemark{k} \\
401  & B  & $1.97\pm0.07$ & $269.60\pm1.29$ & $2.80\pm0.02$ & $4.00\pm2.09$  & 2.68e-02 &      &  Uncertain &  \\
628  & B  & $1.81\pm0.05$ & $310.50\pm1.42$ & $4.11\pm0.28$ & $0.95$  & 6.54e-01 & 2.08 &  Bound & \tablenotemark{l} \\
640  & B  & $0.43\pm0.05$ & $120.30\pm1.90$ & $0.87\pm0.18$ & $0.36\pm0.16$  & 7.15e-03 & 4.44 &  Bound & \tablenotemark{m} \\
652  & B  & $1.22\pm0.05$ & $273.10\pm1.03$ & $1.37\pm0.14$ & $2.25$  & 4.14e-02 & 4.09 &  Bound &  \\
652  & C  & $1.28\pm0.05$ & $274.00\pm1.06$ & $2.55\pm0.09$ & $0.38$  & 6.19e-02 & 3.28 &  Bound &  \\
658  & B  & $1.92\pm0.05$ & $240.60\pm1.67$ & $4.32\pm0.07$ & $3.28$  & 4.83e-01 &      &  Unbound &  \\
697  & B  & $0.66\pm0.05$ & $235.30\pm1.03$ & $0.00\pm0.20$ & $0.00$  & 4.43e-03 & 5.86 &  Bound &  \\
703  & B  & $1.91\pm0.07$ & $33.80\pm1.54$  & $6.82\pm0.54$ & $0.56\pm0.07$  & 4.31e+00 & 2.04 &  Uncertain &  \\
721  & B  & $1.89\pm0.05$ & $195.10\pm1.00$ & $4.62\pm0.27$ & $2.62$  & 1.04e+00 & 2.18 &  Bound &  \\
771  & B  & $1.78\pm0.05$ & $282.40\pm1.06$ & $1.05\pm0.17$ & $0.76\pm0.28$  & 1.80e-01 & 2.00 &  Bound &  \\
841  & B  & $1.97\pm0.05$ & $69.20\pm1.05$  & $4.47\pm0.10$ & $5.45\pm1.73$  & 9.14e-01 &      &  Unbound &  \\
975  & B  & $0.76\pm0.06$ & $130.00\pm1.60$ & $6.18\pm0.10$ & $6.46$  & 5.46e-03 &      &  Unbound & \tablenotemark{n} \\
976  & B  & $0.26\pm0.05$ & $136.10\pm1.13$ & $0.68\pm0.60$ & $0.39$  & 3.28e-04 & 18.64 & Bound &  \\
977  & B  & $0.34\pm0.05$ & $349.10\pm1.23$ & $3.71\pm0.38$ & $1.11$  & 3.70e-03 & 15.16 & Bound &  \\
980  & B  & $0.93\pm0.05$ & $32.40\pm1.00$  & $2.36\pm0.35$ & $1.35$  & 1.02e-02 & 5.46 &  Bound &  \\
984  & B  & $1.74\pm0.06$ & $221.80\pm3.18$ & $0.68\pm0.07$ & $5.06\pm2.17$  & 7.41e-03 &      &  Unbound &  \\
987  & B  & $1.97\pm0.07$ & $225.90\pm1.18$ & $3.82\pm0.08$ & $8.28\pm4.49$  & 1.64e-01 &      &  Unbound & \tablenotemark{o} \\
1119 & B  & $0.54\pm0.07$ & $65.10\pm1.20$  & $4.17\pm0.19$ & $0.65\pm0.42$  & 1.90e-02 & 7.67 &  Bound &  \\
1150 & B  & $0.40\pm0.05$ & $322.50\pm1.93$ & $2.21\pm0.19$ & $1.42\pm0.45$  & 3.07e-03 & 4.36 &  Bound &  \\
1151 & B  & $0.76\pm0.05$ & $307.80\pm1.97$ & $3.85\pm0.19$ & $0.09$  & 2.78e-02 & 3.77 &  Bound &  \\
1174 & B  & $0.62\pm0.07$ & $234.40\pm4.16$ & $6.12\pm0.20$ & $1.91$  & 1.34e-01 & 0.94 &  Bound & \tablenotemark{p} \\
1209 & B  & $0.49\pm0.05$ & $168.40\pm1.02$ & $1.64\pm0.08$ & $5.08$  & 7.63e-03 &      &  Unbound &  \\
1274 & B  & $1.09\pm0.05$ & $242.00\pm1.32$ & $3.98\pm0.10$ & $0.37$  & 3.23e-02 & 2.81 &  Bound &  \\
1299 & B  & $0.88\pm0.05$ & $21.00\pm1.13$  & $6.61\pm0.08$ & $8.09$  & 5.24e-02 &      &  Unbound &  \\
1300 & B  & $0.73\pm0.06$ & $357.80\pm1.04$ & $1.09\pm0.80$ & $9.98$  & 8.02e-02 &      &  Unbound &  \\
1361 & B  & $0.48\pm0.05$ & $312.40\pm1.31$ & $3.63\pm0.09$ & $0.78$  & 3.24e-02 & 2.41 &  Bound & \tablenotemark{q} \\
1375 & B  & $0.77\pm0.06$ & $269.40\pm1.28$ & $3.90\pm0.11$ & $18.95\pm12.83$ & 3.18e-02 &     &  Unbound &  \\
1422 & B  & $0.22\pm0.05$ & $217.30\pm1.15$ & $1.60\pm0.18$ & $1.99\pm1.08$  & 2.51e-03 & 10.06 & Bound &  \\
1463 & B  & $0.22\pm0.07$ & $16.70\pm5.16$  & $3.86\pm0.61$ & $2.90\pm2.80$  & 7.58e-03 &      &  Uncertain &  \\
1531 & B  & $0.37\pm0.05$ & $97.50\pm1.26$  & $2.16\pm0.19$ & $0.38$  & 3.57e-03 & 4.33 &  Bound &  \\
1560 & B  & $1.61\pm0.05$ & $156.90\pm1.00$ & $-0.14\pm0.31$ & $1.28\pm0.32$ & 1.96e-01 & 3.00 &  Bound &  \\
1589 & B  & $0.18\pm0.05$ & $135.50\pm1.68$ & $0.87\pm0.25$ & $0.18$  & 4.91e-03 & 5.16 &  Bound & \tablenotemark{r} \\
1613 & B  & $0.21\pm0.05$ & $184.20\pm1.76$ & $1.25\pm0.09$ & $1.00\pm0.62$  & 7.45e-05 & 6.61 &  Bound &  \\
1677 & B  & $0.60\pm0.05$ & $159.80\pm1.46$ & $4.03\pm0.34$ & $0.39$  & 1.00e-01 & 4.85 &  Bound &  \\
1681 & B  & $0.15\pm0.05$ & $141.30\pm1.12$ & $0.94\pm0.15$ & $3.06\pm1.33$  & 1.51e-03 &      &  Uncertain &  \\
1784 & B  & $0.28\pm0.05$ & $288.40\pm2.43$ & $1.14\pm0.24$ & $0.38\pm0.17$  & 3.13e-03 & 4.97 &  Bound &  \\
1792 & B  & $0.51\pm0.07$ & $301.70\pm4.24$ & $2.38\pm0.15$ & $3.23\pm1.00$  & 3.35e-03 &      &  Uncertain &  \\
1792 & C  & $1.94\pm0.06$ & $109.80\pm2.90$ & $1.25\pm0.11$ & $6.72$  & 1.69e-02 &      &  Unbound &  \\
1830 & B  & $0.46\pm0.05$ & $318.10\pm1.22$ & $2.65\pm0.08$ & $1.13$  & 6.88e-03 & 4.33 &  Bound &  \\
1845 & B  & $1.99\pm0.08$ & $78.50\pm1.42$  & $4.31\pm0.48$ & $7.33\pm0.32$  & 2.42e+00 &      &  Uncertain & \tablenotemark{s} \\
1880 & B  & $1.69\pm0.06$ & $100.00\pm1.18$ & $3.89\pm0.39$ & $14.58\pm0.92$ & 6.25e-01 &      &  Unbound &  \\
1890 & B  & $0.41\pm0.05$ & $144.30\pm1.54$ & $3.20\pm0.12$ & $1.02\pm0.94$  & 3.35e-03 & 6.10 &  Bound &  \\
1908 & B  & $1.26\pm0.05$ & $259.10\pm1.06$ & $3.65\pm0.81$ & $13.26$ & 1.55e+00 &      &  Unbound &  \\
1929 & B  & $1.35\pm0.06$ & $162.90\pm1.00$ & $5.24\pm0.93$ & $3.55$  & 1.04e+00 &      &  Unbound &  \\
1932 & B  & $0.53\pm0.05$ & $115.30\pm1.29$ & $5.06\pm0.41$ & $1.72\pm0.71$  & 1.36e-01 & 5.25 &  Bound &  \\
1962 & B  & $0.12\pm0.05$ & $114.20\pm1.76$ & $0.16\pm0.07$ & $1.60\pm0.77$  & 6.54e-06 & 6.32 &  Bound &  \\
1964 & B  & $0.39\pm0.05$ & $1.70\pm1.15$   & $2.51\pm0.08$ & $4.66\pm1.88$  & 5.30e-04 &      &  Uncertain &  \\
1979 & B  & $0.78\pm0.11$ & $194.00\pm1.72$ & $3.01\pm0.37$ & $30.02\pm3.30$ & 5.29e-02 &      &  Uncertain &  \\
1989 & B  & $0.81\pm0.05$ & $40.40\pm1.36$  & $6.03\pm0.19$ & $0.85$  & 6.85e-02 & 2.11 &  Bound &  \\
2032 & B  & $0.06\pm0.05$ & $128.10\pm1.04$ & $0.32\pm0.27$ & $0.08$  & 4.12e-05 & 10.47 & Bound & \tablenotemark{t} \\
2032 & C  & $1.08\pm0.05$ & $318.60\pm1.37$ & $0.35\pm0.15$ & $6.36\pm2.32$  & 6.68e-03 &      &  Uncertain & \tablenotemark{t} \\
2059 & B  & $0.38\pm0.05$ & $289.70\pm1.15$ & $1.04\pm0.10$ & $2.10\pm1.48$  & 6.32e-04 &      &  Uncertain &  \\
2067 & B  & $1.61\pm0.06$ & $314.40\pm1.33$ & $1.97\pm0.06$ & $7.47\pm3.38$  & 3.34e-02 &      &  Unbound & \tablenotemark{u} \\
2124 & B  & $0.06\pm0.05$ & $54.00\pm1.23$  & $0.26\pm0.12$ & $1.38\pm0.61$  & 4.49e-05 & 6.35 &  Bound &  \\
2159 & B  & $1.98\pm0.06$ & $323.40\pm1.20$ & $5.46\pm0.07$ & $17.98\pm1.32$ & 1.73e-01 &      &  Uncertain &  \\
2174 & B  & $0.86\pm0.05$ & $226.90\pm1.00$ & $0.19\pm0.12$ & $0.46$  & 5.21e-02 & 2.50 &  Bound & \tablenotemark{v} \\
2191 & B  & $1.74\pm0.06$ & $233.20\pm1.11$ & $5.71\pm0.02$ & $4.44\pm1.99$  & 1.02e-01 &      &  Uncertain &  \\
2311 & B  & $1.03\pm0.05$ & $70.20\pm1.16$  & $5.91\pm0.09$ & $5.20\pm2.55$  & 6.95e-02 &      &  Unbound &  \\
2418 & B  & $0.11\pm0.05$ & $3.20\pm1.57$   & $3.11\pm0.15$ & $1.33\pm0.55$  & 2.25e-03 & 5.09 &  Bound & \tablenotemark{w} \\
2463 & B  & $0.66\pm0.08$ & $127.20\pm1.65$ & $0.62\pm0.23$ & $1.14$  & 3.12e-03 & 3.49 &  Bound &  \\
2474 & B  & $0.58\pm0.05$ & $282.50\pm1.00$ & $1.00\pm0.15$ & $0.87$  & 1.07e-02 & 4.91 &  Bound &  \\
2481 & B  & $1.11\pm0.06$ & $183.20\pm1.11$ & $3.71\pm0.04$ & $28.67\pm24.16$ & 5.08e-02 &      & Unbound &  \\
2486 & B  & $0.23\pm0.05$ & $65.30\pm3.47$  & $0.23\pm0.46$ & $1.00$  & 7.14e-04 & 13.07 & Bound &  \\
2626 & B  & $0.16\pm0.05$ & $183.40\pm3.36$ & $1.65\pm0.08$ & $3.37\pm1.74$  & 1.08e-03 &      &  Uncertain &  \\
2626 & C  & $0.21\pm0.05$ & $212.70\pm1.39$ & $0.80\pm0.06$ & $3.36\pm2.67$  & 9.73e-04 &      &  Uncertain &  \\
2657 & B  & $0.76\pm0.07$ & $130.80\pm1.42$ & $0.20\pm0.13$ & $0.34$  & 3.46e-03 & 4.32 &  Bound &  \\
2672 & B  & $0.63\pm0.05$ & $307.60\pm1.63$ & $6.59\pm0.15$ & $3.51\pm1.33$  & 5.41e-02 &      &  Uncertain &  \\
2705 & B  & $1.89\pm0.05$ & $303.60\pm1.56$ & $1.88\pm0.17$ & $0.45$  & 1.59e-01 & 8.13 &  Bound &  \\
2754 & B  & $0.78\pm0.05$ & $260.70\pm1.18$ & $3.09\pm0.23$ & $0.84\pm0.44$  & 1.27e-02 & 3.99 &  Bound & \tablenotemark{x} \\
2837 & B  & $0.35\pm0.05$ & $137.40\pm1.29$ & $0.70\pm0.38$ & $1.19$  & 3.65e-03 & 7.02 &  Bound &  \\
2879 & B  & $0.44\pm0.05$ & $109.80\pm1.60$ & $0.82\pm0.15$ & $4.32\pm1.90$  & 2.99e-03 &      &  Uncertain &  \\
2896 & B  & $0.92\pm0.05$ & $275.90\pm1.09$ & $0.49\pm0.50$ & $1.57$  & 1.79e-02 & 10.92 & Bound &  \\
2904 & B  & $0.69\pm0.05$ & $225.90\pm1.20$ & $3.26\pm0.27$ & $1.96\pm0.85$  & 6.92e-02 & 5.81 &  Bound &  \\
2985 & B  & $1.71\pm0.06$ & $113.00\pm1.45$ & $0.68\pm0.34$ & $0.81\pm0.36$  & 5.22e-01 & 3.92 &  Bound &  \\
3010 & B  & $0.33\pm0.05$ & $304.50\pm1.07$ & $0.41\pm0.17$ & $1.47\pm0.68$  & 6.68e-03 & 6.25 &  Bound &  \\
3020 & B  & $0.38\pm0.05$ & $272.70\pm1.61$ & $2.53\pm0.40$ & $0.26$  & 2.10e-02 & 7.29 &  Bound & \tablenotemark{y} \\
3042 & B  & $1.86\pm0.05$ & $146.90\pm1.08$ & $1.71\pm0.82$ & $3.63\pm0.65$  & 2.37e+00 &      &  Unbound &  \\
3049 & B  & $0.47\pm0.05$ & $196.90\pm3.54$ & $0.50\pm0.14$ & $0.91$  & 3.71e-03 & 5.73 &  Bound & \tablenotemark{z} \\
3112 & B  & $1.77\pm0.06$ & $152.50\pm1.15$ & $0.87\pm0.81$ & $5.25\pm1.53$  & 9.41e-01 &      &  Unbound &  \\
3156 & B  & $1.32\pm0.05$ & $202.40\pm1.06$ & $2.00\pm0.59$ & $0.68$  & 5.06e-03 & 10.68 & Bound &  \\
3158 & B  & $1.84\pm0.05$ & $252.90\pm1.09$ & $3.03\pm0.02$ & $2.49\pm1.31$  & 1.21e-03 &      &  Uncertain &  \\
3168 & B  & $0.81\pm0.05$ & $332.50\pm1.01$ & $5.18\pm0.09$ & $1.28\pm0.74$  & 1.10e-02 & 6.80 &  Bound &  \\
3214 & B  & $0.48\pm0.05$ & $317.50\pm1.00$ & $1.67\pm0.86$ & $1.34$  & 3.52e-02 & 16.63 & Bound & \tablenotemark{aa} \\
3234 & B  & $0.06\pm0.05$ & $123.20\pm1.02$ & $0.77\pm0.26$ & $0.09$  & 2.44e-05 & 9.94 &  Bound &  \\
3245 & B  & $1.52\pm0.05$ & $184.70\pm1.06$ & $4.77\pm0.15$ & $5.44\pm1.82$  & 6.97e-02 &      &  Unbound &  \\
3255 & B  & $0.18\pm0.05$ & $336.60\pm1.32$ & $0.40\pm0.10$ & $0.92\pm0.43$  & 6.46e-04 & 5.79 &  Bound & \tablenotemark{bb} \\
3263 & B  & $0.82\pm0.05$ & $274.90\pm1.14$ & $2.13\pm0.01$ & $7.88$  & 3.30e-03 &      &  Unbound &  \\
3284 & B  & $0.44\pm0.05$ & $193.10\pm1.01$ & $3.50\pm0.15$ & $5.03$  & 9.71e-03 &      &  Unbound & \tablenotemark{cc} \\
3349 & B  & $0.31\pm0.05$ & $265.90\pm1.00$ & $2.83\pm0.36$ & $1.78$  & 3.76e-02 & 6.67 &  Bound & \tablenotemark{dd} \\
3444 & B  & $1.08\pm0.05$ & $9.60\pm1.06$   & $2.94\pm0.14$ & $1.23\pm0.76$  & 1.52e-01 & 3.86 &  Bound & \tablenotemark{ee} \\
3456 & B  & $0.04\pm0.05$ & $16.50\pm1.04$  & $0.05\pm0.16$ & $0.82$  & 1.62e-05 & 7.85 &  Bound &  \\
3471 & B  & $0.53\pm0.05$ & $229.00\pm1.01$ & $3.74\pm1.31$ & $2.16$  & 2.69e-01 & 13.87 & Bound &  \\
3579 & B  & $1.70\pm0.05$ & $180.00\pm1.73$ & $0.40\pm0.54$ & $1.04\pm0.87$  & 4.68e-01 & 6.07 &  Bound &  \\
3870 & B  & $1.78\pm0.07$ & $172.40\pm1.02$ & $2.87\pm0.86$ & $10.56$ & 7.89e+00 &      &  Unbound & \tablenotemark{ff} \\
3907 & B  & $1.58\pm0.05$ & $163.10\pm1.15$ & $5.45\pm0.55$ & $1.15\pm0.70$  & 2.06e+00 & 6.24 &  Bound & \tablenotemark{gg} \\
4004 & B  & $1.92\pm0.06$ & $218.50\pm1.01$ & $5.32\pm0.13$ & $3.98$  & 2.18e-01 &      &  Unbound &  \\
4013 & B  & $0.93\pm0.05$ & $62.50\pm1.04$  & $1.22\pm0.00$ & $7.27\pm3.44$  & 1.65e-05 &      &  Unbound &  \\
4021 & B  & $1.75\pm0.06$ & $113.90\pm1.31$ & $0.22\pm0.47$ & $0.80\pm0.83$  & 7.28e-02 & 7.49 &  Bound &  \\
4033 & B  & $1.62\pm0.05$ & $110.60\pm1.00$ & $2.98\pm0.02$ & $6.88$  & 5.99e-03 &      &  Unbound & \tablenotemark{hh} \\
4149 & B  & $1.62\pm0.05$ & $64.90\pm1.00$  & $-0.11\pm0.50$ & $1.54\pm0.41$ & 2.05e-01 & 5.16 &  Bound &  \\
4203 & B  & $1.15\pm0.05$ & $41.80\pm1.05$  & $0.71\pm0.08$ & $2.80\pm1.02$  & 8.18e-04 &      &  Uncertain &  \\
4273 & B  & $0.18\pm0.05$ & $354.30\pm1.01$ & $2.23\pm0.36$ & $0.03$  & 1.76e-03 & 8.42 &  Bound &  \\
4287 & B  & $0.58\pm0.05$ & $79.10\pm1.06$  & $1.65\pm0.33$ & $0.55\pm0.25$  & 3.15e-03 & 10.02 & Bound &  \\
4329 & B  & $1.85\pm0.05$ & $118.40\pm1.00$ & $4.98\pm0.24$ & $0.68$  & 2.10e-01 & 1.95 &  Bound &  \\
4359 & B  & $0.87\pm0.05$ & $108.40\pm1.63$ & $6.70\pm0.46$ & $0.98\pm0.69$  & 4.22e-01 & 4.73 &  Bound &  \\
4399 & B  & $0.11\pm0.05$ & $346.60\pm2.81$ & $1.57\pm0.16$ & $0.13$  & 5.87e-05 & 9.23 &  Bound & \tablenotemark{ii} \\
4458 & B  & $1.50\pm0.05$ & $61.10\pm1.03$  & $4.84\pm0.32$ & $2.56$  & 1.74e+00 & 3.44 &  Bound & \tablenotemark{jj} \\
4550 & B  & $1.05\pm0.08$ & $143.40\pm1.09$ & $0.94\pm0.12$ & $1.89\pm0.86$  & 8.76e-02 & 5.23 &  Bound &  \\
4986 & B  & $0.17\pm0.05$ & $312.60\pm1.01$ & $-0.05\pm0.08$ & $12.35$ & 3.53e-04 &      & Unbound &  \\
5211 & B  & $0.08\pm0.05$ & $90.60\pm1.41$  & $2.37\pm0.24$ & $2.48$  & 2.08e-04 & 11.70 & Bound &  \\
5361 & B  & $1.82\pm0.05$ & $200.80\pm1.00$ & $1.10\pm0.42$ & $1.78\pm1.08$  & 3.13e-01 & 5.17 &  Bound &  \\
5545 & B  & $0.08\pm0.05$ & $284.00\pm1.08$ & $0.08\pm0.09$ & $3.29\pm1.61$  & 5.55e-05 &      &  Uncertain &  \\
5570 & B  & $0.13\pm0.05$ & $211.40\pm1.85$ & $0.63\pm0.25$ & $2.44$  & 7.03e-04 & 9.68 &  Bound & \tablenotemark{kk} \\
5578 & B  & $0.32\pm0.05$ & $97.90\pm1.12$  & $0.92\pm0.08$ & $4.92\pm1.59$  & 1.00e-04 &      &  Uncertain &  \\
5618 & B  & $1.38\pm0.05$ & $51.70\pm1.26$  & $0.39\pm0.80$ & $2.00\pm1.05$  & 5.86e-01 &      &  Uncertain &  \\
5654 & B  & $1.74\pm0.05$ & $150.90\pm1.01$ & $1.36\pm0.23$ & $0.62$  & 2.10e-01 & 4.12 &  Bound &  \\
5736 & B  & $0.85\pm0.05$ & $304.40\pm1.00$ & $2.74\pm0.30$ & $2.65$  & 3.90e-02 & 4.62 &  Bound &  \\
5822 & B  & $0.41\pm0.05$ & $201.80\pm1.31$ & $0.17\pm0.07$ & $1.26\pm1.70$  & 4.22e-04 &      &  Uncertain &  \\
5845 & B  & $1.65\pm0.05$ & $29.30\pm1.61$  & $0.01\pm0.42$ & $0.50\pm0.20$  & 9.38e-03 & 5.28 &  Bound &  \\
5949 & B  & $0.69\pm0.05$ & $257.10\pm1.09$ & $4.99\pm0.38$ & $1.15$  & 9.10e-02 & 3.67 &  Bound &  \\
5971 & B  & $0.04\pm0.05$ & $132.60\pm2.67$ & \nodata       & $4.01$  & \nodata  &      &  Unbound &  \\
6109 & B  & $0.52\pm0.05$ & $323.90\pm1.07$ & $2.65\pm0.50$ & $1.43\pm0.69$  & 1.55e-02 & 10.10 & Bound &  \\
6203 & B  & $1.93\pm0.05$ & $60.20\pm1.02$  & $0.27\pm0.31$ & $1.10\pm0.11$  & 1.70e-01 & 2.90 &  Bound &  \\
6380 & B  & $1.66\pm0.05$ & $194.80\pm1.02$ & $2.04\pm0.81$ & $4.83$  & 2.74e+00 &      &  Unbound &  \\
6425 & B  & $1.83\pm0.05$ & $245.30\pm1.04$ & $5.76\pm1.07$ & $0.37$  & 1.09e+01 & 8.09 &  Uncertain &  \\
6450 & B  & $0.82\pm0.05$ & $320.70\pm1.04$ & $4.28\pm0.10$ & $4.79$  & 5.98e-02 &      &  Unbound &  \\
6482 & B  & $0.48\pm0.05$ & $273.60\pm1.00$ & $1.00\pm0.44$ & $1.73$  & 3.53e-02 & 7.74 &  Bound & \tablenotemark{ll} \\
6676 & B  & $0.94\pm0.05$ & $32.70\pm1.02$  & $6.87\pm0.36$ & $1.43$  & 9.72e-01 & 2.03 &  Bound &  \\
6895 & B  & $1.78\pm0.05$ & $153.30\pm1.00$ & $0.08\pm0.80$ & $2.17\pm1.82$  & 5.06e-01 &      &  Uncertain &  \\
7235 & B  & $0.11\pm0.05$ & $277.80\pm3.74$ & $0.06\pm0.26$ & $1.39$  & 2.71e-04 & 8.40 &  Bound &  \\
7448 & B  & $0.86\pm0.05$ & $260.60\pm1.00$ & $1.99\pm0.91$ & $0.37$  & 5.11e-02 & 11.26 & Bound & \tablenotemark{mm} \\
7455 & B  & $1.77\pm0.05$ & $305.90\pm1.01$ & $3.27\pm0.07$ & $3.23$  & 1.64e-02 &      &  Unbound & \tablenotemark{nn} \\
7470 & B  & $1.48\pm0.06$ & $301.10\pm1.33$ & $0.16\pm0.97$ & $1.62\pm1.15$  & 2.78e-01 & 10.99 & Bound &  \\
7554 & B  & $1.57\pm0.05$ & $255.40\pm1.01$ & $5.12\pm0.08$ & $8.02$  & 8.61e-01 &      &  Unbound &  \\
7587 & B  & $0.40\pm0.05$ & $180.00\pm1.00$ & $3.08\pm0.09$ & $16.23$ & 1.82e-03 &      &  Unbound &  \\
\enddata
\tablecomments{Tabular summary of companions to KOI host stars in our sample. Columns (1)--(4) describe the positions of the companions, taken from F17. Column (5) describes the \dm\ in the \eke bandpass between the primary and companion, taken from F17 for uncertain and unbound companions, and from our isochrone models for the bound companions. Columns (6)--(9) describe the physical association assessment, described in~\autoref{physical_association}. In brief, the Color Offset (6) indicates the difference, in units of the model and measurement uncertainty $\sigma$, between the modeled and observed colors of the companion. Better agreement means the companion is more likely to be physically bound. The error quoted for color offset is the standard deviation of the color offset, which describes the variation in the color offset among several different filter pairs. Systems with only a single color measurement have no reported standard deviation. Systems with poor agreement have higher color offset standard deviation, and are likely to be reclassified as uncertain. Background probability (7) and bound probability (8) refer to the analysis using TRILEGAL galaxy models \cite{Girardi2005} and solar neighborhood stellar multiplicity statistics \citep{Raghavan2010} respectively, in~\autoref{trilegal}. Notes on additional companions to individual systems, marked with letters in Column (10), can be found in~\autoref{table_refs}.
}
\end{deluxetable}

\clearpage
\begin{deluxetable}{cccc|ccccc}
\tabletypesize{\textwidth}
\tablecolumns{9}
\tablewidth{0pt}
\tabletypesize{\scriptsize}
\tablecaption{Radius Correction Factors
\label{table:rad_corrs}}
\tablehead{
\colhead{KOI} & \colhead{Companions} & \colhead{$\Delta Kp_{21}$} & \colhead{$\Delta Kp_{31}$} & \colhead{KOI} & \colhead{$R_{p}$(Assumed} & \colhead{$R_p$ Reference} & \colhead{$X_{R_1}$} & \colhead{$X_{R_2}$} \\
& \colhead{within 2''} & \colhead{(mag)} & \colhead{(mag)} & \colhead{(Planet)} & \colhead{Single)} & \colhead{(Confirmed Planets)} & & }
\startdata
5 & 2 & 0.29 & 3.01 & 5.01 & 7.07 &  & 1.352 & 2.239 \\
 &  &  &  & 5.02 & 0.66 &  & 1.352 & 2.239 \\
105 & 1 & 3.75 &  & 105.01 & $3.05 \pm 0.43$ & \cite{Morton2016} & 1.016 & \nodata \\
112 & 1 & 1.04 &  & 112.01 & $2.85 \pm0.22$ & \cite{Morton2016} & 1.176 & \nodata \\
 &  &  &  & 112.02 & $1.25 \pm 0.1$ & \cite{Morton2016} & 1.176 & \nodata \\
118 & 1 & 4.73 &  & 118.01 & $2.25 \pm 0.3$ & \cite{Morton2016} & 1.006 & \nodata \\
119 & 1 & 0.29 &  & 119.01 & $8.65 \pm 0.48$ & \cite{Rowe2014} & 1.330 & 1.347 \\
 &  &  &  & 119.02 & $8.18 \pm 0.51$ & \cite{Rowe2014} & 1.330 & 1.347 \\
120 & 1 & 2.96 &  & 120.01 & 31.09 &  & 1.032 & \nodata \\
141 & 1 & 1.00 &  & 141.01 & 5.14 &  & 1.182 & \nodata \\
174 & 1 & 3.79 &  & 174.01 & $2.36\pm0.37$ & \cite{Morton2016} & 1.015 & 2.673 \\
177 & 1 & 0.81 &  & 177.01 & 1.94 &  & 1.214 & 1.431 \\
191 & 1 & 2.90 &  & 191.01 & $11.42\pm0.69$ & \cite{Morton2016} & 1.034 & \nodata \\
 &  &  &  & 191.02 & 2.79 &  & 1.034 & \nodata \\
 &  &  &  & 191.03 & 1.25 &  & 1.034 & \nodata \\
 &  &  &  & 191.04 & $2.68\pm0.49$ & \cite{Morton2016} & 1.034 & \nodata \\
227 & 1 & 1.24 &  & 227.01 & 2.45 &  & 1.149 & 1.568 \\
268 & 1 & 4.24 &  & 268.01 & 3.03 &  & 1.010 & \nodata \\
270 & 1 & -0.21 &  & 270.01 & $2.056 \pm 0.069$ & \cite{VanEylen2015} & 1.488 & \nodata \\
 &  &  &  & 270.02 & $2.764 \pm 0.086$ & \cite{VanEylen2015} & 1.488 & \nodata \\
279 & 1 & 2.94 &  & 279.01 & $6.14\pm0.33$ & \cite{VanEylen2015} & 1.033 & \nodata \\
 &  &  &  & 279.02 & $2.62\pm0.14$ & \cite{VanEylen2015} & 1.033 & \nodata \\
 &  &  &  & 279.03 & $0.837\pm0.068$ & \cite{VanEylen2015} & 1.033 & \nodata \\
285 & 1 & 6.23 &  & 285.01 & $3.51\pm0.10$ & \cite{Xie2014} & 1.002 & 4.479 \\
 &  &  &  & 285.02 & $2.60\pm0.08$ & \cite{Xie2014} & 1.002 & 4.479 \\
 &  &  &  & 285.03 & $2.067\pm0.056$ & \cite{VanEylen2015} & 1.002 & 4.479 \\
294 & 1 & 4.91 &  & 294.01 & $2.69\pm0.49$ & \cite{Morton2016} & 1.005 & \nodata \\
298 & 1 & 0.68 &  & 298.01 & $1.46\pm0.06$ & \cite{Morton2016} & 1.239 & \nodata \\
 &  &  &  & 298.02 & 1.68 &  & 1.239 & \nodata \\
307 & 1 & 0.10 &  & 307.01 & $1.66\pm0.28$ & \cite{Morton2016} & 1.382 & 1.418 \\
 &  &  &  & 307.02 & $1.07\pm0.07$ & \cite{Morton2016} & 1.382 & 1.418 \\
356 & 1 & 2.70 &  & 356.01 & $3.10\pm0.3$ & \cite{Morton2016} & 1.041 & 1.281 \\
364 & 1 & 1.42 &  & 364.01 & 0.93 &  & 1.127 & 1.406 \\
379 & 2 & 2.28 & 1.03 & 379.02 & 1.97 &  & 1.229 & 1.747 \\
387 & 3 & 7.25 & 2.81 & 387.01 & $2.5\pm0.26$ & \cite{Morton2016} & 1.038 & \nodata \\
401 & 1 & 2.80 &  & 401.01 & $4.21\pm0.60$ & \cite{Rowe2014} & 1.037 & \nodata \\
 &  &  &  & 401.02 & $3.96\pm0.68$ & \cite{Rowe2014} & 1.037 & \nodata \\
 &  &  &  & 401.03 & $1.61\pm0.24$ & \cite{Rowe2014} & 1.037 & \nodata \\
628 & 1 & 4.11 &  & 628.01 & $2.29\pm0.28$ & \cite{Morton2016} & 1.011 & 3.201 \\
640 & 2 & 0.87 & 7.26 & 640.01 & $2.21\pm0.33$ & \cite{Morton2016} & 1.205 & 1.667 \\
652 & 2 & 1.37 & 2.55 & 652.01 & $4.45\pm0.14$ & \cite{Morton2016} & 1.174 & 1.816 \\
658 & 1 & 4.32 &  & 658.01 & $2.57\pm0.52$ & \cite{Rowe2014} & 1.009 & \nodata \\
 &  &  &  & 658.02 & $2.47\pm0.47$ & \cite{Rowe2014} & 1.009 & \nodata \\
 &  &  &  & 658.03 & $1.44\pm0.28$ & \cite{Rowe2014} & 1.009 & \nodata \\
697 & 1 & 0.00 &  & 697.01 & $2.96\pm0.51$ & \cite{Morton2016} & 1.414 & 1.414 \\
721 & 1 & 4.62 &  & 721.01 & $2.07\pm0.37$ & \cite{Morton2016} & 1.007 & 3.492 \\
771 & 1 & 1.05 &  & 771.01 & 14.41 &  & 1.175 & 1.559 \\
841 & 1 & 4.47 &  & 841.01 & $4.85\pm0.18$ & \cite{Morton2016} & 1.008 & \nodata \\
 &  &  &  & 841.02 & $7.17\pm0.18$ & \cite{Morton2016} & 1.008 & \nodata \\
 &  &  &  & 841.03 & 2.83 &  & 1.008 & \nodata \\
 &  &  &  & 841.04 & 3.27 &  & 1.008 & \nodata \\
984 & 1 & 0.68 &  & 984.01 & 2.78 &  & 1.239 & \nodata \\
987 & 1 & 3.82 &  & 987.01 & $1.21\pm0.05$ & \cite{Morton2016} & 1.015 & \nodata \\
1119 & 1 & 4.17 &  & 1119.02 & 0.66 &  & 1.011 & 3.345 \\
1150 & 1 & 2.21 &  & 1150.01 & $0.89\pm0.09$ & \cite{Morton2016} & 1.063 & 1.937 \\
1151 & 1 & 3.85 &  & 1151.01 & $0.89\pm0.16$ & \cite{Rowe2014} & 1.014 & 3.083 \\
 &  &  &  & 1151.02 & $1.0\pm0.18$ & \cite{Rowe2014} & 1.014 & 3.083 \\
 &  &  &  & 1151.03 & $0.66\pm0.05$ & \cite{Morton2016} & 1.014 & 3.083 \\
 &  &  &  & 1151.04 & 0.81 &  & 1.014 & 3.083 \\
 &  &  &  & 1151.05 & 0.88 &  & 1.014 & 3.083 \\
1174 & 1 & 6.12 &  & 1174.01 & 2.99 &  & 1.002 & 3.084 \\
1209 & 1 & 1.64 &  & 1209.01 & 5.95 &  & 1.105 & \nodata \\
1300 & 1 & 1.09 &  & 1300.01 & $1.46\pm0.06$ & \cite{Morton2016} & 1.169 & \nodata \\
1375 & 1 & 3.90 &  & 1375.01 & 5.06 &  & 1.014 & \nodata \\
1531 & 1 & 2.16 &  & 1531.01 & $1.23\pm0.05$ & \cite{Morton2016} & 1.066 & 2.057 \\
1560 & 1 & -0.14 &  & 1560.01 & 13.97 &  & 1.463 & 1.443 \\
1589 & 1 & 0.87 &  & 1589.01 & $2.23\pm0.10$ & \cite{Xie2013a} & 1.204 & 1.388 \\
 &  &  &  & 1589.02 & $2.36\pm0.11$ & \cite{Xie2013a} & 1.204 & 1.388 \\
 &  &  &  & 1589.03 & $2.60\pm0.49$ & \cite{Rowe2014} & 1.204 & 1.388 \\
 &  &  &  & 1589.04 & $1.38\pm0.27$ & \cite{Rowe2014} & 1.204 & 1.388 \\
 &  &  &  & 1589.05 & $2.20\pm0.43$ & \cite{Rowe2014} & 1.204 & 1.388 \\
1613 & 1 & 1.25 &  & 1613.01 & $1.2\pm0.06$ & \cite{Morton2016} & 1.147 & 1.342 \\
 &  &  &  & 1613.02 & 1.02 &  & 1.147 & 1.342 \\
1677 & 1 & 4.03 &  & 1677.01 & $2.31\pm0.57$ & \cite{Morton2016} & 1.012 & 3.261 \\
 &  &  &  & 1677.02 & 0.81 &  & 1.012 & 3.261 \\
1681 & 1 & 0.94 &  & 1681.01 & 0.99 &  & 1.192 & \nodata \\
 &  &  &  & 1681.02 & 0.69 &  & 1.192 & \nodata \\
 &  &  &  & 1681.03 & 0.71 &  & 1.192 & \nodata \\
 &  &  &  & 1681.04 & 0.77 &  & 1.192 & \nodata \\
1784 & 1 & 1.14 &  & 1784.01 & 7.55 &  & 1.162 & 1.505 \\
1792 & 2 & 2.38 & 1.25 & 1792.01 & $4.21\pm0.37$ & \cite{Morton2016} & 1.195 & \nodata \\
 &  &  &  & 1792.03 & $1.19\pm0.11$ & \cite{Morton2016} & 1.195 & \nodata \\
1830 & 1 & 2.65 &  & 1830.01 & $2.35\pm0.1$ & \cite{Morton2016} & 1.042 & 2.409 \\
 &  &  &  & 1830.02 & $3.65\pm0.15$ & \cite{Morton2016} & 1.042 & 2.409 \\
1845 & 1 & 4.31 &  & 1845.01 & $1.48\pm0.09$ & \cite{Morton2016} & 1.009 & \nodata \\
 &  &  &  & 1845.02 & 5.99 &  & 1.009 & \nodata \\
1880 & 1 & 3.89 &  & 1880.01 & $1.6\pm0.27$ & \cite{Morton2016} & 1.014 & \nodata \\
1890 & 1 & 3.20 &  & 1890.01 & $1.71\pm0.06$ & \cite{Morton2016} & 1.026 & 2.023 \\
1908 & 1 & 3.65 &  & 1908.01 & $1.32\pm0.16$ & \cite{Rowe2014} & 1.017 & \nodata \\
 &  &  &  & 1908.02 & $1.11\pm0.20$ & \cite{Rowe2014} & 1.017 & \nodata \\
1929 & 1 & 5.24 &  & 1929.01 & $2.05\pm0.57$ & \cite{Rowe2014} & 1.004 & \nodata \\
 &  &  &  & 1929.02 & $1.54\pm0.43$ & \cite{Rowe2014} & 1.004 & \nodata \\
1932 & 1 & 5.06 &  & 1932.01 & $3.37\pm0.77$ & \cite{Rowe2014} & 1.005 & 3.291 \\
 &  &  &  & 1932.02 & $2.53\pm0.60$ & \cite{Rowe2014} & 1.005 & 3.291 \\
1962 & 1 & 0.16 &  & 1962.01 & 2.28 &  & 1.366 & 1.371 \\
1964 & 1 & 2.51 &  & 1964.01 & 0.72 &  & 1.048 & \nodata \\
1979 & 1 & 3.01 &  & 1979.01 & $1.17\pm0.28$ & \cite{Morton2016} & 1.031 & \nodata \\
1989 & 1 & 6.03 &  & 1989.01 & $2.08\pm0.18$ & \cite{Morton2016} & 1.002 & 4.469 \\
2032 & 3 & 0.32 & 0.35 & 2032.01 & $1.49\pm0.4$ & \cite{Morton2016} & 1.793 & 1.940 \\
2059 & 1 & 1.04 &  & 2059.01 & $0.8\pm0.04$ & \cite{Morton2016} & 1.176 & \nodata \\
 &  &  &  & 2059.02 & 0.51 &  & 1.176 & \nodata \\
2067 & 1 & 1.97 &  & 2067.01 & $1.63\pm0.33$ & \cite{Morton2016} & 1.079 & \nodata \\
2124 & 1 & 0.26 &  & 2124.01 & 1.00 &  & 1.338 & 1.440 \\
2159 & 1 & 5.46 &  & 2159.01 & 1.18 &  & 1.003 & \nodata \\
2174 & 1 & 0.19 &  & 2174.01 & 1.65 &  & 1.355 & 1.440 \\
 &  &  &  & 2174.02 & 1.88 &  & 1.355 & 1.440 \\
 &  &  &  & 2174.03 & 1.24 &  & 1.355 & 1.440 \\
2191 & 1 & 5.71 &  & 2191.01 & 1.01 &  & 1.003 & \nodata \\
2311 & 1 & 5.91 &  & 2311.01 & 0.95 &  & 1.002 & \nodata \\
 &  &  &  & 2311.02 & 0.77 &  & 1.002 & \nodata \\
 &  &  &  & 2311.03 & $1.33\pm0.08$ & \cite{Morton2016} & 1.002 & \nodata \\
2418 & 1 & 3.11 &  & 2418.01 & $1.4\pm0.13$ & \cite{Morton2016} & 1.028 & 1.509 \\
2463 & 1 & 0.62 &  & 2463.01 & $1.18\pm0.2$ & \cite{Morton2016} & 1.251 & 1.419 \\
2481 & 1 & 3.71 &  & 2481.01 & 15.96 &  & 1.016 & \nodata \\
2486 & 1 & 0.23 &  & 2486.01 & 1.43 &  & 1.345 & 1.381 \\
2626 & 2 & 1.65 & 0.80 & 2626.01 & 1.12 &  & 1.303 & \nodata \\
2657 & 1 & 0.20 &  & 2657.01 & 0.52 &  & 1.355 & 1.423 \\
2672 & 1 & 6.59 &  & 2672.01 & $5.30 \pm1.95$ & \cite{Xie2014} & 1.001 & \nodata \\
 &  &  &  & 2672.02 & $3.50\pm1.28$ & \cite{Xie2014} & 1.001 & \nodata \\
2705 & 1 & 1.88 &  & 2705.01 & $1.34\pm0.72$ & \cite{Morton2016} & 1.085 & 1.277 \\
2754 & 1 & 3.09 &  & 2754.01 & $0.71\pm0.06$ & \cite{Morton2016} & 1.029 & 2.576 \\
2837 & 1 & 0.70 &  & 2837.01 & $2.15\pm0.8$ & \cite{Morton2016} & 1.234 & 1.373 \\
2904 & 1 & 3.26 &  & 2904.01 & $1.88\pm0.39$ & \cite{Morton2016} & 1.025 & 1.943 \\
3010 & 1 & 0.41 &  & 3010.01 & $1.78\pm0.14$ & \cite{Morton2016} & 1.299 & 1.444 \\
3020 & 1 & 2.53 &  & 3020.01 & 1.15 &  & 1.048 & 1.852 \\
3042 & 1 & 1.71 &  & 3042.01 & 1.95 &  & 1.098 & \nodata \\
3049 & 4 & 0.50 & 5.82 & 3049.01 & $1.66\pm0.14$ & \cite{Morton2016} & 1.283 & 1.511 \\
3112 & 1 & 0.87 &  & 3112.01 & 1.51 &  & 1.204 & \nodata \\
3156 & 1 & 2.00 &  & 3156.01 & 4.77 &  & 1.076 & 1.356 \\
 &  &  &  & 3156.04 & 33.62 &  & 1.076 & 1.356 \\
3168 & 1 & 5.18 &  & 3168.01 & 1.16 &  & 1.004 & 3.297 \\
3234 & 1 & 0.77 &  & 3234.01 & $0.97\pm0.22$ & \cite{Morton2016} & 1.221 & 1.391 \\
3245 & 1 & 4.77 &  & 3245.01 & $0.96\pm0.2$ & \cite{Morton2016} & 1.006 & \nodata \\
3263 & 1 & 2.13 &  & 3263.01 & 6.89 &  & 1.068 & \nodata \\
3349 & 1 & 2.83 &  & 3349.01 & $2.92\pm0.99$ & \cite{Morton2016} & 1.036 & 2.365 \\
3444 & 1 & 2.94 &  & 3444.01 & 0.82 &  & 1.033 & 1.795 \\
 &  &  &  & 3444.03 & 0.55 &  & 1.033 & 1.795 \\
 &  &  &  & 3444.04 & 0.82 &  & 1.033 & 1.795 \\
3456 & 1 & 0.05 &  & 3456.01 & $0.93\pm0.05$ & \cite{Morton2016} & 1.398 & 1.426 \\
 &  &  &  & 3456.02 & 1.18 &  & 1.398 & 1.426 \\
4004 & 1 & 5.32 &  & 4004.01 & 2.24 &  & 1.004 & \nodata \\
4021 & 1 & 0.22 &  & 4021.01 & 2.04 &  & 1.347 & 1.352 \\
 &  &  &  & 4021.02 & 1.63 &  & 1.347 & 1.352 \\
4149 & 1 & -0.11 &  & 4149.01 & 1.61 &  & 1.450 & 1.440 \\
 &  &  &  & 4149.02 & 1.62 &  & 1.450 & 1.440 \\
4273 & 1 & 2.23 &  & 4273.01 & 0.97 &  & 1.062 & 1.560 \\
4287 & 1 & 1.65 &  & 4287.01 & 0.98 &  & 1.104 & 1.311 \\
 &  &  &  & 4287.02 & 0.76 &  & 1.104 & 1.311 \\
4329 & 1 & 4.98 &  & 4329.01 & 0.71 &  & 1.005 & 4.054 \\
4359 & 1 & 6.70 &  & 4359.01 & 0.79 &  & 1.001 & 4.749 \\
4399 & 1 & 1.57 &  & 4399.01 & 2.77 &  & 1.111 & 1.827 \\
4458 & 1 & 4.84 &  & 4458.01 & 2.48 &  & 1.006 & 3.674 \\
4550 & 1 & 0.94 &  & 4550.01 & 1.73 &  & 1.191 & 1.626 \\
4986 & 1 & -0.05 &  & 4986.01 & 1.48 &  & 1.431 & \nodata \\
5211 & 1 & 2.37 &  & 5211.01 & 0.57 &  & 1.055 & 2.245 \\
5545 & 1 & 0.08 &  & 5545.01 & 1.08 &  & 1.389 & \nodata \\
5570 & 1 & 0.63 &  & 5570.01 & 2.28 &  & 1.249 & 1.493 \\
5578 & 1 & 0.92 &  & 5578.01 & 3.00 &  & 1.195 & \nodata \\
5736 & 1 & 2.74 &  & 5736.01 & 1.34 &  & 1.039 & 2.464 \\
5822 & 1 & 0.17 &  & 5822.01 & 1.69 &  & 1.361 & \nodata \\
5949 & 1 & 4.99 &  & 5949.01 & 1.88 &  & 1.005 & 3.894 \\
6425 & 1 & 5.76 &  & 6425.01 & 1.50 &  & 1.002 & \nodata \\
6450 & 1 & 4.28 &  & 6450.01 & 0.78 &  & 1.010 & \nodata \\
6676 & 1 & 6.87 &  & 6676.01 & 1.81 &  & 1.001 & 5.060 \\
7235 & 1 & 0.06 &  & 7235.01 & 1.15 &  & 1.396 & 1.420 \\
7455 & 1 & 3.27 &  & 7455.01 & 2.44 &  & 1.024 & \nodata \\
7470 & 1 & 0.16 &  & 7470.01 & 1.90 &  & 1.364 & 1.417 \\
7554 & 1 & 5.12 &  & 7554.01 & 1.98 &  & 1.004 & \nodata \\
7587 & 1 & 3.08 &  & 7587.01 & 2.19 &  & 1.029 & \nodata \\
\enddata
\tablecomments{Radius corrections for planets and planet candidates in the stellar KOI sample. Column (1) indicates the KOI system. Column (2) gives the number of companions detected within 2''. Columns (3) and (4) provide the $\Delta Kp$ values for the secondary ($\Delta Kp_{21}$) and tertiary companion ($\Delta Kp_{31}$). Column (5) indicates the planetary candidate or confirmed planet. Column (6) gives the previous best estimate of the planet radius, either from the \eke pipeline for planet candidates, or from the Exoplanet Archive\footnote{http://exoplanetarchive.ipac.caltech.edu/index.html} (with literature source in Column (7)) for the confirmed planets. Columns (8) and (9) provide our calculated radius correction factors, assuming the planets orbit the primary ($X_{R_1}$) or secondary  ($X_{R_2}$) star. In these columns, \nodata refers to systems in which we have too little information to calculate the radius correction factor because the system is uncertain or unbound, and we have no accurate estimate of the companion's stellar radius. \\
Three target KOI hosts have more than 2 companions within 2'', which we include in the dilution correction analysis. These are KOI 387 ($\Delta Kp_3 = 7.58$ mag), KOI 2032 ($\Delta Kp_3 = 0.32$ mag), and KOI 3049 ($\Delta Kp_3 = 7.48$ mag, $\Delta Kp_4 = 4.92$ mag). \\
Only one target, KOI 652, has 2 bound companions within 2'', allowing calculation of $X_{R_3}$ in this system. KOI 652 has $X_{R_3} = 2.657$.
}
\end{deluxetable}
\clearpage

\appendix
\section{Notes from Table 1}
\label{table_refs}
$^{a}$Additional companion to KOI 5 detected in \kbande by \cite{Kraus2016} at 0.03'', but no color information is available to perform physical association assessment.

$^{b}$Two additional companions detected around KOI 68; both are separated by more than 2'', so are not included in the sample. KOI 68 C is located at 2.74'' and appears to be unbound. KOI 68 D is located at 3.41'' and appears to be bound.

$^{c}$Additional companion to KOI 105 detected at 3.88'' has uncertain designation.

$^{d}$Three additional companions are detected near KOI 113. KOI 113 C was detected in \kbande at 1.32'' by \cite{Kraus2016}, but no color information is available. Two additional companions at 3.21 and 3.63'' are outside the separation limit of our sample. KOI 113 D appears bound, while KOI 113 E appears unbound.

$^{e}$Two additional companions are detected near KOI 177, at 2.29 and 2.87'' respectively. Neither of these companions has available color information.

$^{f}$Additional companion to KOI 227 detected at 3.35'' in \kband, but no color information available.

$^{g}$Additional companion to KOI 268 detected at 2.51'' appears to be unbound.

$^{h}$Additional companion to KOI 285 detected at 2.33'' appears to be bound.

$^{i}$Additional companion to KOI 298 detected at 2.01'' in $i$-band by AstraLux, but no color information is available.

$^{j}$Additional companion to KOI 378 detected at 2.91'' appears to be bound.

$^{k}$Two additional companions to KOI 387 detected. Inner companion, KOI 387 B, is located at 0.65'' and outer companion, KOI 387 D, is located at 1.9''. Both are detected in \kband by \cite{Kraus2016} but lack color information.

$^{l}$Additional companion to KOI 628 located at 2.75'' has uncertain designation.

$^{m}$Additional companion to KOI 640 was detected at 1.17'' in \kbande by \cite{Kraus2016}, but lacks color information.

$^{n}$Additional companion to KOI 975 detected at 1.08'' in \kbande. No color information is available.

$^{o}$Addtional companions to KOI 987 was detected at 2.19''. No color information available.

$^{p}$Two additional companions to KOI 1174 detected at 3.09 and 3.48''. No color information available for either companion.

$^{q}$Additional companion to KOI 1361 detected at 1.1'' in \kband, but no color information available.

$^{r}$Additional companion to KOI 1589 detected at 3.94'' appears to be bound.

$^{s}$Additional companion to KOI 1845 detected at 2.89'' appears to be unbound.

$^{t}$Additional companion to KOI 2032 detected at 1.15'' in \kbande at Palomar/PHARO. No color information available for this companion.

$^{u}$Additional companion detected to KOI 2067 at 2.76''. No color information available for this companion.

$^{v}$Additional companion to KOI 2174 detected at 3.85'' appears to be bound.

$^{w}$Two additional companions to KOI 2418 detected, at separations of 2.39 and 3.92''. KOI 2418 C was detected in \kband, but has no color information available. KOI 2418 D was detected in \jbande and \kband, and has uncertain designation.

$^{x}$Additional companion to KOI 2754 detected at 2.43''. No color information available for this companion.

$^{y}$Additional companion to KOI 3020 detected at 3.77''. No color information available for this companion.

$^{z}$Four additional companions to KOI 3049 detected. KOI 3049 C, D, and E are located at 0.76'', 1.21'', and 1.22'' but no color information is available. KOI 3049 F is detected at 2.95'' by HST and in \hband, and appears to be unbound.

$^{aa}$Additional companion to KOI 3214 detected at 1.26'' in \hbande, but not color information is available for this companion.

$^{bb}$Additional companion to KOI 3255 detected at 3.0'' appears to be unbound.

$^{cc}$Additional companion to KOI 3284 detected at 3.96'' appears to be unbound.

$^{dd}$Additional companion to KOI 3349 detected at 3.12'' appears to be unbound.

$^{ee}$Four additional companions to KOI 3444 detected. KOI 3444 C, D, and E are located at 2.04'', 3.06'', and 3.39'' respectively, but no color information is available for these companions. KOI 3444 F is located at 3.53'', and appears to be unbound.

$^{ff}$Additional companion to KOI 3870 detected at 3.11'' appears to be unbound.

$^{gg}$Additional companion to KOI 3907 detected at 2.77'' appears to be bound.

$^{hh}$Additional companion to KOI 4033 detected at 2.92'' appears to be bound.

$^{ii}$Three additional companions to KOI 4399 detected at 2.12'', 2.33'', and 2.68''. No color information is available for any of these companions.

$^{jj}$Additional companion to KOI 4458 detected at 2.04'' appears to be unbound.

$^{kk}$Additional companion to KOI 5570 detected at 2.04'' appears to be unbound.

$^{ll}$Additional companion to KOI 6482 detected at 3.55'' appears to be unbound.

$^{mm}$Additional companion to KOI 7448 detected at 3.74'' appears to be unbound.

$^{nn}$Additional companion to KOI 7455 detected at 3.29'' appears to be unbound.


\end{document}